\documentclass[aps,pra,amsmath,twocolumn,amssymb,superscriptaddress]{revtex4-1}

\usepackage{amssymb,amsfonts,amsmath}

\usepackage{dsfont}
\usepackage{color}
\usepackage[apple mac]{inputenc}
\usepackage[english]{babel}
\usepackage{latexsym}
\usepackage{fancyhdr}
\usepackage{verbatim}
\usepackage{graphicx}
\usepackage{epsf}
\usepackage{subfigure}
\usepackage{bm}
\usepackage{epstopdf}\usepackage[colorlinks=true , citecolor=blue,urlcolor=blue]{hyperref}
\usepackage{units}
 
 \def\bs#1{\boldsymbol{#1}}

\usepackage[
top    = 1.5cm,
bottom = 1.50cm,
left   = 1.5cm,
right  = 0.9cm]{geometry}

\definecolor{Nathanblue}{rgb}{0.96,0.24,0.00}

\definecolor{Nathanbluetitle}{rgb}{0.,0.24,0.51}

\definecolor{orange}{rgb}{0.96,0.24,0.00}

\def\be{\begin{equation}}
\def\ee{\end{equation}}
\def\bs#1{\boldsymbol{#1}}

\newcommand{\e}{\mathrm{e}}
\newcommand{\DAB}{\Delta_{\text{AB}}}

\begin{document}

\title{Creating topological interfaces and detecting chiral edge modes in a 2D optical lattice}

\author{N. Goldman}
%\email[]{nathan.goldman@lkb.ens.fr}
\email[]{ngoldman@ulb.ac.be}
%\affiliation{Laboratoire Kastler Brossel, Coll\`ege de France, CNRS, ENS-PSL Research University, UPMC-Sorbonne Universit\'es, 11 place Marcelin Berthelot, 75005, Paris, France}
\affiliation{CENOLI, Facult{\'e} des Sciences, Universit{\'e} Libre de Bruxelles (U.L.B.), B-1050 Brussels, Belgium}

\author{G. Jotzu}
\affiliation{Institute for Quantum Electronics, ETH Zurich, 8093 Zurich, 
Switzerland}
\author{M. Messer}
\affiliation{Institute for Quantum Electronics, ETH Zurich, 8093 Zurich, 
Switzerland}
\author{F. G\"org}
\affiliation{Institute for Quantum Electronics, ETH Zurich, 8093 Zurich, 
Switzerland}
\author{R. Desbuquois}
\affiliation{Institute for Quantum Electronics, ETH Zurich, 8093 Zurich, 
Switzerland}
\author{T. Esslinger}
\affiliation{Institute for Quantum Electronics, ETH Zurich, 8093 Zurich, 
Switzerland}

\date{\today}

\begin{abstract}
We propose and analyze a general scheme to create chiral topological edge modes within the bulk of two-dimensional engineered quantum systems. Our method is based on the implementation of topological interfaces, designed within the bulk of the system, where topologically-protected edge modes localize and freely propagate in a unidirectional manner. This scheme is illustrated through an optical-lattice realization of the Haldane model for cold atoms~\cite{Jotzu2014}, where an additional spatially-varying lattice potential induces distinct topological phases in separated regions of space. We present two realistic experimental configurations, which lead to linear and radial-symmetric topological interfaces, which both allows one to significantly reduce the effects of external confinement on topological edge properties. Furthermore, the versatility of our method opens the possibility of tuning the position, the localization length and the chirality of the edge modes, through simple adjustments of the lattice potentials. In order to demonstrate the unique detectability offered by engineered interfaces, we numerically investigate the time-evolution of wave packets, indicating how topological transport unambiguously manifests itself within the lattice. Finally, we analyze the effects of disorder on the dynamics of chiral and non-chiral states present in the system. Interestingly, engineered disorder is shown to provide a powerful tool for the detection of topological edge modes in cold-atom setups. 
\end{abstract}

\maketitle

\section{Introduction}

%\subsection{General introduction}

Historically, the discovery of the quantum Hall (QH) effect revealed two major concepts that revolutionized our knowledge of quantum transport~\cite{vonKlitzing:1980,Yoshioka}: the remarkable quantization of the Hall conductivity in terms of topological invariants~\cite{Thouless:1982,Niu:1985,Haldane1988}, and the simultaneous existence of robust unidirectional (chiral) modes that propagate along the edge of the system. These  topological transport properties, which are intimately connected through the bulk-edge correspondence~\cite{Halperin:1982,Rammel:1983,MacDonald:1984,Hatsugai:1993PRL,Hatsugai:1993PRB,Qi:2006theorem}, recently found their counterparts in a wide family of quantum systems: the topological insulators, superconductors and superfluids~\cite{Hasan:2010,Qi2011,Bernevig:2013}.

Detecting and analyzing the properties of topological edge excitations constitutes an intense field of research since the early days of the QH effect~\cite{AllenJr:1983,Glattli:1985,Mast:1985,Wen:1990,Johnson:1991,Ashoori:1992,Meir:1994,Wen:1995,Haldane:1995,Kane:1994,Kane:1995,Milliken:1996,Ji:2003,Bid:2010,Venkatachalam:2012,Gurman:2012,Inoue:2013,Goldstein:2016}. The chiral nature of QH edge modes was first revealed through edge-magnetoplasma experiments~\cite{Ashoori:1992} (see also the pioneer measurements reported in Refs.~\cite{AllenJr:1983,Glattli:1985,Mast:1985}), while the topological order associated with fractional quantum Hall (FQH) edge modes~\cite{Wen:1990,Moon:1993} was first detected by measuring tunneling currents between distinct edges~\cite{Milliken:1996}. A striking demonstration of the outline trajectory performed by QH edge states was provided by a double-slit-electron-interferometer experiment, which realized a QH-based Mach-Zehnder interferometer~\cite{Ji:2003}. Interestingly, such geometries have been considered to probe the fractional (anyonic) statistics of FQH excitations~\cite{Goldstein:2016}. More recent experiments also revealed signatures of exotic counter-propagating (``neutral upstream") modes in FQH liquids~\cite{Bid:2010,Gurman:2012,Venkatachalam:2012,Inoue:2013}, in agreement with early theoretical works~\cite{Haldane:1995,Wen:1995,Kane:1994,Kane:1995}.

Spatially-resolved edge currents were also detected in two-dimensional (2D) topological insulators~\cite{Nowack2013, Konig2013,Yang2012}, via charge transport measurements and scanning tunneling microscopy, offering an instructive view on the quantum spin Hall effect~\cite{Hasan:2010, Qi2011}. Similar techniques were also exploited to observe spatially-resolved (non-chiral) edge currents in graphene and graphene nanoribbons \cite{Tao2011, Allen2015}. Furthermore, topological surface states (``2D Dirac fermions") were observed in 3D topological insulators using angle-resolved photoemission spectroscopy (ARPES) \cite{Hsieh2008, Hsieh2009, Zhang2009, Xia2009}.

Today,  basic concepts of QH systems and topological insulators are well established, both in theory and through experimental measurements. However, intriguing and more obscure aspects of these topological phases of matter~\cite{Hasan:2010, Qi2011} could be further explored and exploited using the controllability of engineered quantum systems. In this context, ultra-cold atoms in optical lattices can offer a promising route towards the realization of (potentially exotic) topological phases, through the implementation of well-designed Hamiltonians leading to distinct topological orders~\cite{Cooper:2008,Dalibard:2011,Goldman:2014,Goldman:2015BEC}. 

 Recent experiments successfully achieved to load cold atomic gases into 2D Bloch bands with non-trivial topological properties~\cite{Aidelsburger2013, Jotzu2014,Aidelsburger2015}, using the notion of Floquet engineering~\cite{Oka2009,Kitagawa:2010,Lindner:2011,Kolovsky:2011,Bermudez:2011,Cayssol:2013,Goldman2014a, Zheng:2014,Bukov2015}: in this cold-atom context, this consists in trapping a gas in an optical lattice and to subject the system to a (high-frequency) time-periodic modulation.  In general, Floquet engineering has been implemented in optical lattices by directly shaking the
lattice potential~\cite{Lignier:2007,Sias:2008,Eckardt:2009,Zenesini:2009,Struck:2011,Arimondo:2012,Struck2012,Struck:2013h,Jotzu2014}, or by including additional ``moving" optical lattices~\cite{Aidelsburger2013, Miyake2013,Aidelsburger2015,Kennedy:2015}, or time-dependent external fields~\cite{Jimenez-Garcia2012,Luo:2015,Jotzu:2015}. Such driven 2D optical-lattice settings were used to probe various manifestations of the Berry curvature~\cite{Duca2015, Flaschner2015}, including the anomalous (transverse) velocity in response to an applied force~\cite{Jotzu2014,Aidelsburger2015}; recent experiments also reported on the measurement of non-zero Chern numbers~\cite{Aidelsburger2015,Wu:2015}.

 Bulk QH properties have been revealed in recent cold-atom experiments~\cite{LeBlanc:2012,Jotzu2014,Aidelsburger2015,Wu:2015}, and preliminary results on the identification of chiral edge modes include the observation of unidirectional motion in ladder geometries (``QH stripes"), where cold atoms were subjected to a synthetic magnetic flux~\cite{Atala2014,Mancini2015, Stuhl2015}. These experiments were performed in two-leg ladders created by optical potentials~\cite{Atala2014}, but also, in three-leg ladders~\cite{Mancini2015, Stuhl2015} built on the concept of synthetic dimensions (i.e.~the three legs of the ladders were associated with three internal states of an atom~\cite{Celi:2014}).  In addition, a very recent work~\cite{Leder2016} has reported on the observation of a point-like edge state, situated at the interface between two geometrically distinct regions, in a one-dimensional optical lattice reminiscent of the  Su-Schrieffer-Heeger model~\cite{Su:1979}. Finally, we point out that topological structures were also identified in other engineered systems, such as photonic lattices \cite{Rechtsman2013, Mittal:2014,Lu2014,Gao2016,Mukherjee:2016,Maczewsky:2016}, superconducting qubits \cite{Schroer2014, Roushan2014}, mechanical systems~\cite{Susstrunk:2015} and radio-frequency circuits \cite{Hu2015, Ningyuan2015}.

\begin{figure}[h!]
\includegraphics[width=9.3cm]{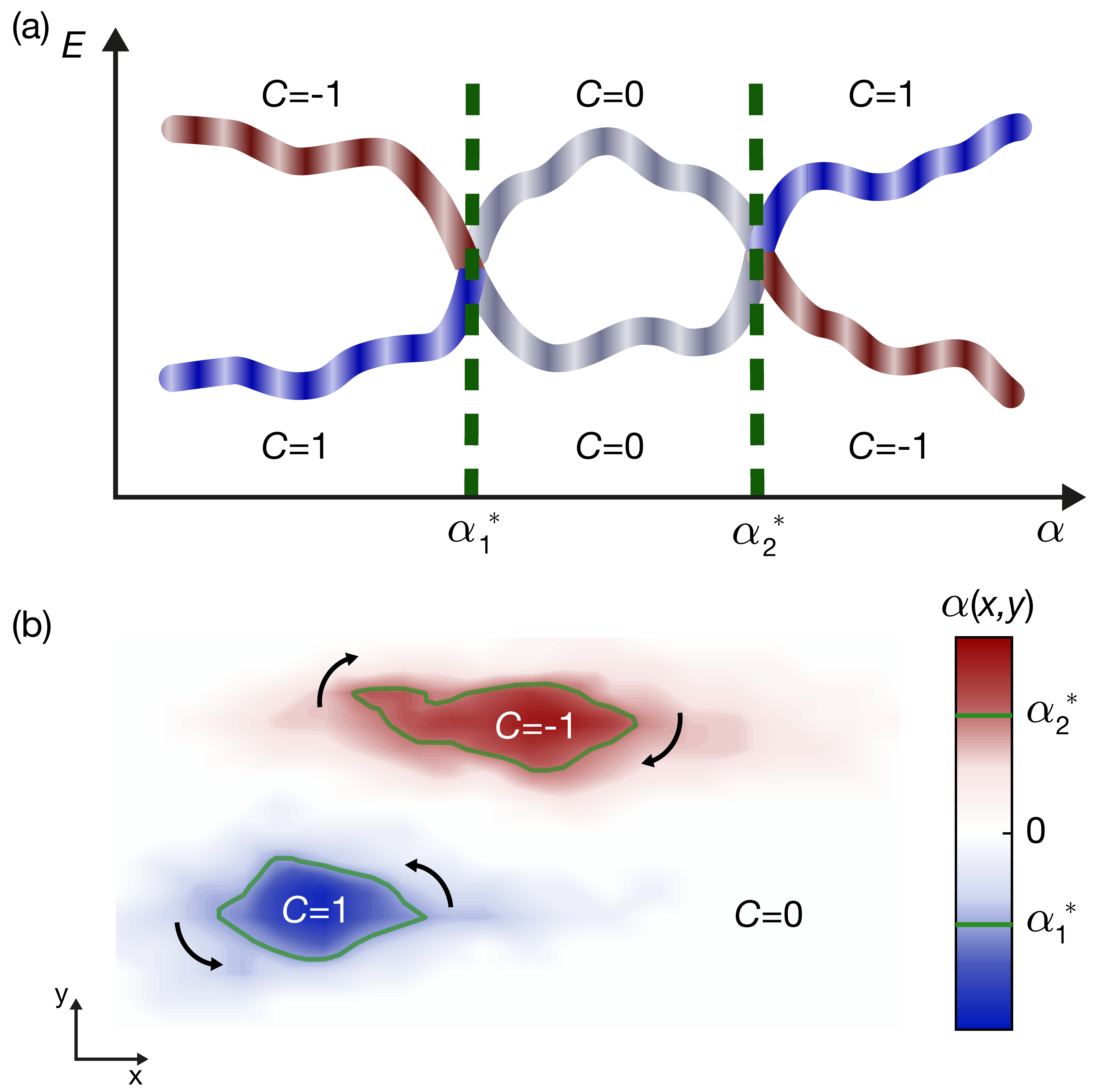}
\vspace{-0.cm} \caption{Illustration of topological interfaces in a two-band model.  
(a) Abstract two-band model realizing different topological phases. 
Depending on the value of a controllable parameter $\alpha$, the two bulk bands are either topologically trivial (Chern number $C\!=\!0$) or non-trivial ($C\!=\!\pm 1$), where $\alpha^{\ast}_{\mathrm{1}}$ and $\alpha^{\ast}_{\mathrm{2}}$ indicate the transition points (i.e.~gap-closing points).  
(b) Spatially-varying the parameter $\alpha (x,y)$ in the 2D plane generates different regions, which are associated with distinct (spatially-resolved) topological phases. In this figure, each region is labeled by the Chern number $C$ of the lowest bulk band [see panel (a)], which is evaluated locally in space. The singular spatial regions where $\alpha (x,y)\!=\!\alpha^{\ast}_{\mathrm{1, 2}}$ define the topological interfaces within the 2D plane, where topologically-protected ``edge" modes are located and propagate.}
\label{Fig_topology}
\end{figure}

In 2D systems, topological interfaces consist of boundary lines separating two distinct topologically-ordered regions, where topologically-protected ``edge" modes are located and propagate [see Fig.~\ref{Fig_topology} and Section~\ref{sect:interfaces_general}]. In this work, we introduce a scheme realizing topological interfaces within a 2D optical lattice, which offers the unique possibility of probing, manipulating and tuning the properties of topological edge modes in ultracold atomic gases. Such controllable properties include the location, the localization length, the chirality, and the trajectory of the propagating topological modes. Our proposal is based on the recent realization of the Haldane model~\cite{Jotzu2014}, which uses ultracold fermions on a honeycomb optical lattice (see also Refs.~\cite{Oka2009,Zheng:2014}). As further described below in Section~\ref{sect:interfaces_general}, this scheme offers an ideal platform to investigate edge-state physics \emph{within the bulk} of a cold atomic gas~\cite{GoldmanPRL2010,Tenenbaum:2013,Reichl:2014}, hence limiting the effects of external confinement. In particular, the corresponding topological edge modes appear at genuine topological phase transitions (which, in principle, can be associated with arbitrary changes in the Chern number of the bands), and without the simultaneous action of a potential step. This proposal opens an exciting avenue for the exploration of topological edge modes belonging to various topological classes~\cite{Hasan:2010,Qi2011,Bernevig:2013}, in a highly controllable environment.

\begin{figure}[h!]
\includegraphics[width=9.3cm]{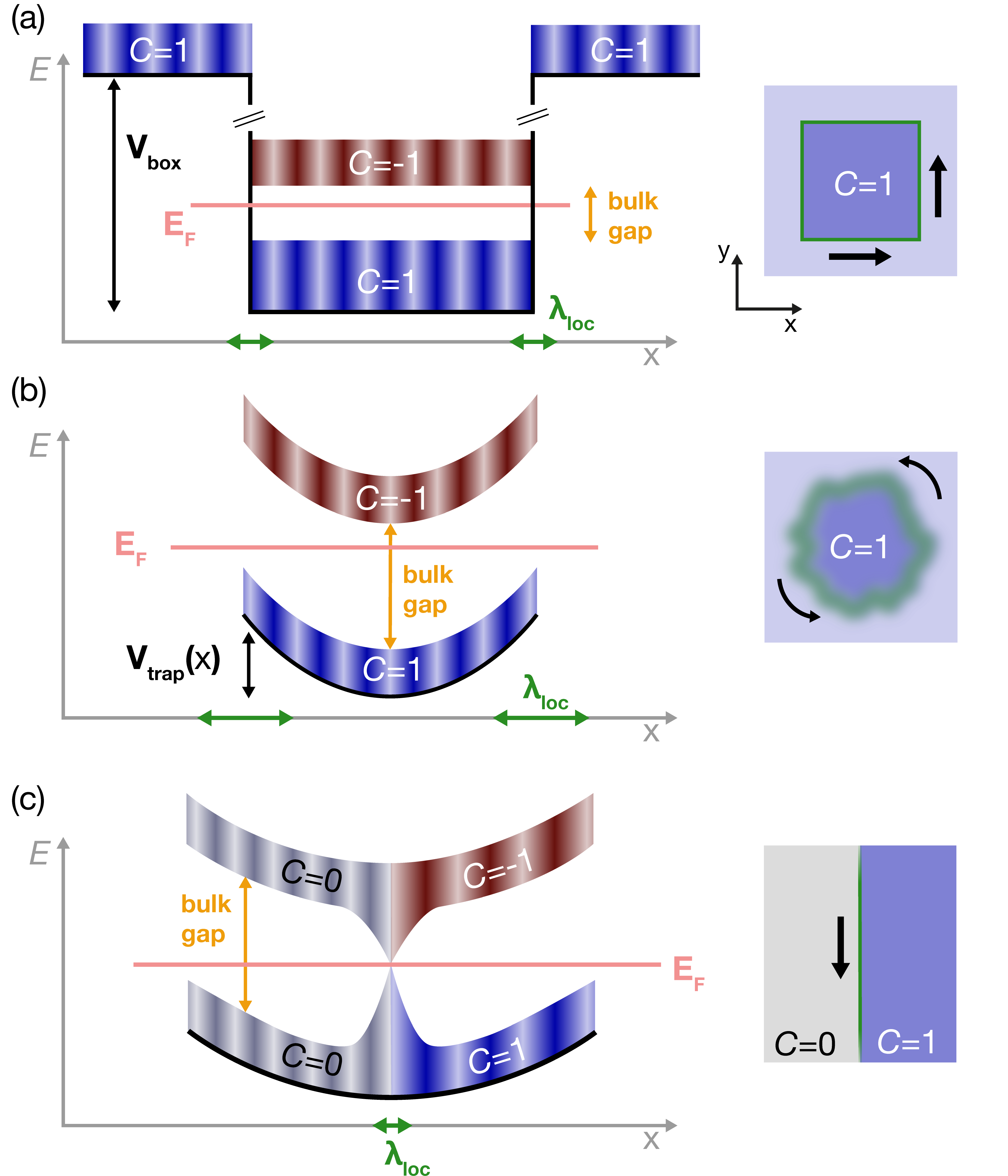}
\vspace{-0.cm} \caption{Edge-state structures in various external potentials and interfaces, for an abstract two-band model. \emph{Left of each panel}: Schematics of the bulk energy bands $E_{\pm}(\bs x)$, as evaluated locally in space along the $x$ direction. The Chern numbers $C$ associated with the bulk bands are indicated; the Fermi energy $E_{\text{F}}$ is set within the bulk gap, where edge modes (not shown) are expected whenever the bands reach the topological regimes $C\!=\!\pm1$. The location and localization length $\lambda_{\text{loc}}$ of the edge modes is represented on the $x$ axis. The external potential or confinement is indicated by thick black lines. \emph{Right of each panel}: Illustration of the corresponding topological edge modes in the 2D plane. The chirality and the localization length of the edge modes are indicated, as well as the Chern number of the lowest band (note that vacuum is associated with $C\!=\!0$). (a) In an ideal box-like potential $V_{\mathrm{box}}$, the bulk bands are shifted to higher energies. This naturally generates a topological interface between the inner topological system ($C\!=\!1$) and vacuum ($C\!=\!0$). In this standard case, chiral edge modes appear at the topological interfaces defined by the edges of the box. (b) In the case of a smoothly varying  potential (e.g.~a harmonic trap), the bulk bands are locally deformed, and edge modes potentially survive within the bulk gaps. However, smooth traps significantly increase the localization length and reduce the group velocity of the edge modes. (c) Varying a system parameter in space allows for the creation of tunable topological interfaces within the 2D plane [Fig.~\ref{Fig_topology}]: topological edge modes are located in the vicinity of the transition point, which can be designed at the center of the (smooth) trap. In the latter case, robust edge modes appear at a genuine topological phase transition (here, $C\!=\!0\!\leftrightarrow\!1$), without the simultaneous action of a potential step [panel (a)].}
\label{Fig_traps}\end{figure}

\subsection{Engineered interfaces and the detection of topologically-protected modes}\label{sect:interfaces_general}

In the standard realization of the QH effect in solids, topological edge modes appear at the physical boundary of a two-dimensional electron gas~\cite{Halperin:1982}, which is typically set by the confining potential created by an external metallic electrode gate~\cite{Yoshioka,Chklovskii:1992}. While theoretical models generally assume that QH systems display sharp edges, the confining potentials of real QH samples are in fact quite smooth: the electronic density slowly drops to zero in the vicinity of the edge of the electron gas~\cite{Chklovskii:1992,Dempsey:1993,Chamon:1994,Wen:1995}. From a theoretical point of view, the smooth nature of the confining potential was also shown to generate additional edge-state dispersion branches in the spectrum, as compared to the ideal sharp-edges configuration, see Refs.~\cite{Chamon:1994,Meir:1994,StanescuPRA,CocksPRA}.

In cold-atom experiments, the atomic cloud is generally confined by an external optical harmonic (or quartic) potential. As was discussed in Refs.~\cite{CocksPRA,GoldmanPRL,GoldmanEPJST,GoldmanPNAS}, this smooth confinement can significantly affect the properties of topological edge states. In cold-atom systems realizing the QH effect, chiral propagating modes were shown to survive in the presence of smooth external traps, however, their localization length was found to be largely increased and their velocity significantly reduced~\cite{GoldmanPRL,GoldmanEPJST,GoldmanPNAS}. Furthermore, the distinction between bulk states and edge states becomes complicated in the limit of a purely harmonic trap~\cite{CocksPRA}. Altogether, this strongly limits the prospect of probing and analyzing topological-edge-state physics in current cold-atom experiments, suggesting the necessity of developing methods to design sharp (box) confinement for these systems~\cite{Gaunt:2013}. The distinction between the box-potential and smooth-confinement configurations is illustrated in Fig.~\ref{Fig_traps}(a)-(b), where real-space spectra and topological edge states are schematically represented.

Importantly, the bulk-edge correspondence emanating from topological band theory is not limited to physical edges, which are defined as the boundary separating a sample (e.g.~an electron gas or a cold-atom gas) from vacuum~\cite{Hasan:2010}. Indeed, the general bulk-edge correspondence states that any \emph{interface} separating two topologically-different regions of space necessarily hosts topologically-protected edge modes. While this includes the standard case of a sample surrounded by vacuum (whose topology is trivial), this suggests the intriguing possibility of engineering topological interfaces \emph{within} a sample~\cite{GoldmanPRL2010,Tenenbaum:2013,Reichl:2014}, e.g.~in a region of space where the effects of external confinement are strongly limited [Fig.~\ref{Fig_traps}(c)]. Moreover, in a QH system exhibiting chiral edge states, engineered interfaces would offer a tool to design flexible guides for the propagation of topologically-protected modes. 

Since topological interfaces play a central role in this work, let us briefly describe this notion using a simple local-density-approximation (LDA) argument. Let us consider an abstract two-band model depending on a constant parameter $\alpha$, which is topologically trivial for $\alpha^{\ast}_{\mathrm{1}} \!<\!\alpha\!<\! \alpha^{\ast}_{\mathrm{2}}$ and topologically non-trivial otherwise [see Fig. \ref{Fig_topology}(a)]. In the non-trivial regime, the bulk gap hosts topologically-protected edge states, localized at the physical boundary of the system, in agreement with the bulk-edge correspondence. Now, let us suppose that the parameter $\alpha(\bs x)$ can be varied continuously in space. Then, in an LDA approach, one can estimate the band structure locally in space, in a region located around some position $\bs x^*$, based on the value $\alpha(\bs x^*)$ that the spatially-varying parameter takes there. Following the topological phase diagram of the model, one finds that different regions of the lattice can then be associated with different topological phases [Fig.~\ref{Fig_topology}(b)]: the bulk bands that are evaluated locally can have zero or non-zero topological invariants depending on the value $\alpha(\bs x)$. In particular, some singular regions are associated with a (local) gapless band structure: this occurs when $\alpha(\bs x)\!=\!\alpha^{\ast}_{\mathrm{1,2}}$, which defines the local topological interfaces \emph{within} the lattice. The band structure being locally gapless at these interfaces, and since the latter are associated with a change in the topology, these regions host topologically-protected modes~\cite{Qi:2006theorem,Hasan:2010}. These modes share the general properties of the topological edge-states associated with the uniform model, except that they are now localized within the interior of the system [Fig. \ref{Fig_topology}(b) and Fig.~\ref{Fig_traps}(c)].

The heart of our proposal is to create a tunable topological interface at the center of a two-dimensional ultracold gas, through a suitable adjustment of optical-lattice parameters. As schematically represented in Fig.~\ref{Fig_traps}(c), our scheme creates different topological regions within an optical lattice, and is designed so as to localize topologically-protected modes at the center of the trap. This configuration offers a promising platform for the study of topologically-protected modes in cold-atom experiments, where the effects associated with inter-particle interactions~\cite{Chin:2010} and disorder~\cite{Sanchez-Palencia/Lewenstein} could be analyzed in a clean and controllable way. We focus our study on a two-band system realizing the QH effect, hence exhibiting unidirectional (chiral) topological modes, and discuss possible extensions in Section~\ref{Section:conclusions}. Importantly, the versatility of our scheme allows one to design the shape of the interface within the 2D optical lattice, to control its location, but also, to tune the localization length of the associated topological propagating states. Finally, we point out that the number of topological modes (dispersion branches) associated with an interface is directly given by $n_{\text{int}}\!=\!\vert C_1 \!-\! C_2 \vert$, where $C_{1,2}$ are the Chern numbers of the lowest bulk band evaluated in the two spatial regions separated by the interface~\cite{Hasan:2010}; these topological invariants, and hence the number of modes, could also be tuned in a cold-atom experiment.

\subsection{Outline}

Our paper is structured as follows. In Sec. \ref{section:Haldanemodel} we summarize and present the Haldane model and its general topological properties. 
We recapitulate the main features and motivate our choice of the Haldane model that lays out the basis for creating and manipulating topological interfaces. 
In Sec. \ref{sect:space_dep} we propose a new and variable method to create and probe topological interfaces in cold-atom experiments. 
We discuss the general strategy of generating a topological interface in the center of the system, by spatially varying the lattice potential, and study the corresponding edge-state structures. 
In Sec. \ref{section:radial_interface} we advance our idea of spatially differing topological phases to a radial geometry, and discuss how we can realize a radial-symmetric topological interface. 
Sec. \ref{Section:dynamics} is dedicated to the actual measurement, and provides numerical calculations for possible observables of the topological edge mode appearing at the interface. 
We present wave-packet dynamics for our proposed schemes, both for projections onto edge and bulk states, and show how this allows one to probe the motion of chiral edge modes in the presence of a harmonic trap. 
In Sec. \ref{Section:disorderb}, we study how the dynamics are affected by the presence of disorder in the lattice. In particular, we show that disorder can be used to improve the detection of the chiral propagating modes, by reducing the dispersion of non-chiral (bulk) states.
The spatially differing optical lattice configurations and their possible realization in a realistic experimental setup is described in Sec. \ref{Section:experiment}.  
We conclude and summarize our tunable approach of generating topological interfaces in Sec. \ref{Section:conclusions}, and explore further possible applications and outlooks. 

%%%%%%%%%%%%%%%%%%%%%%%%%%%%%%%%%%%%%%%%%%%%%%%%%%%%%%

\section{The Haldane model \\ and the topological phase diagram}\label{section:Haldanemodel}

Before introducing our proposal for the creation of topological interfaces, let us briefly summarize the general topological properties of the model considered in this work.  This will allow us to introduce the relevant parameters of the model and the notations used in the following. The choice of this specific model will be motivated at the end of this section (see \ref{section:motivations}).

\subsection{The model}\label{section:themodel}

\begin{figure}[h!]
\includegraphics[width=9.3cm]{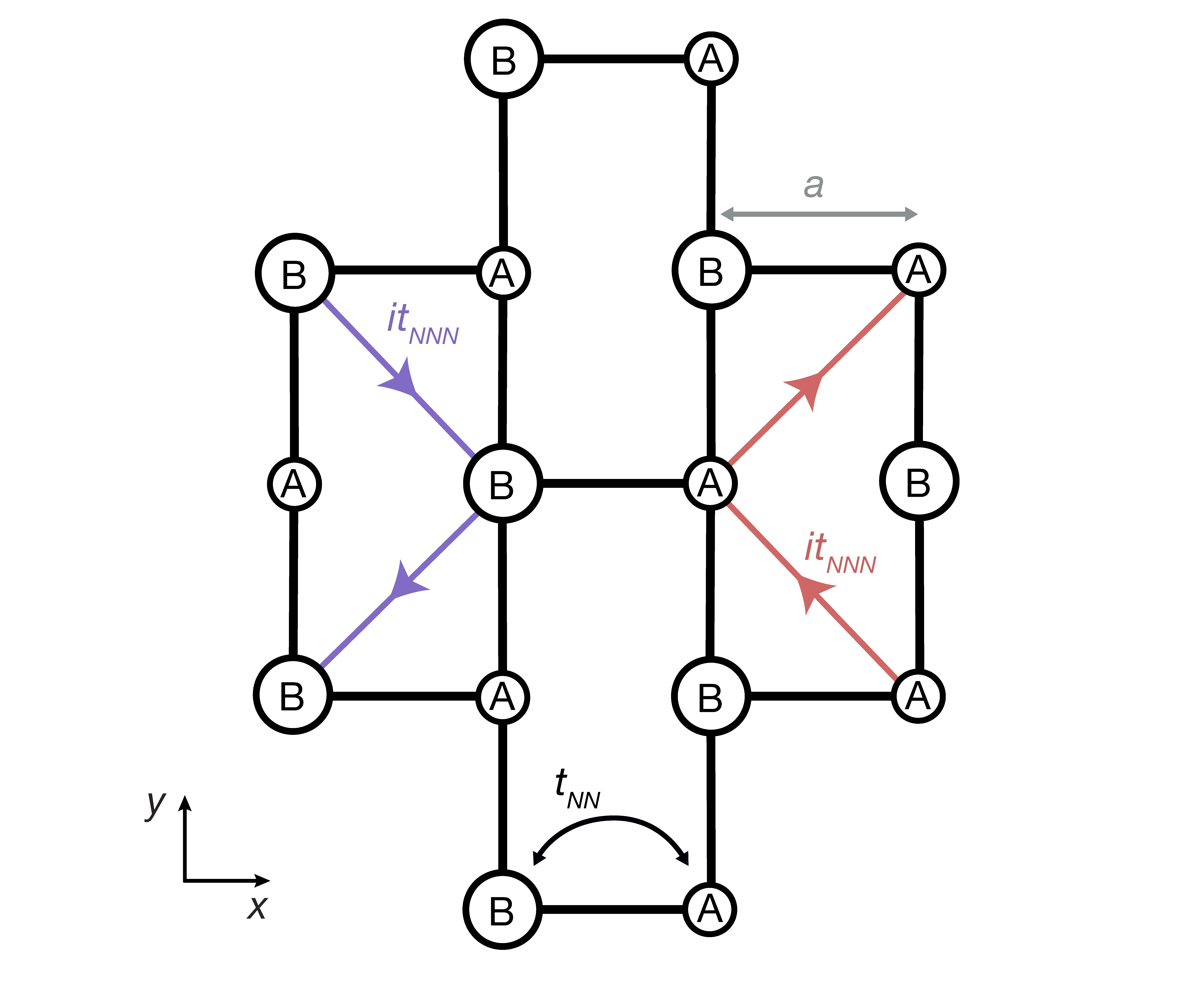}
\vspace{-0.cm} \caption{The Haldane model defined on the ``brickwall" lattice used in this work. The lattice spacing between NN sites is denoted $a$, the NN [resp.~NNN] tunneling amplitude is denoted  $t_{\text{NN}}$ [resp.~$t_{\text{NNN}}$]. The NNN tunneling matrix elements $ \pm i t_{\text{NNN}}$ are complex; their sign is positive [resp.~negative] for clockwise [resp.~anti-clockwise] paths. Non-equivalent lattice sites are denoted $A$ and $B$. Note that we neglect purely vertical NNN hoppings, which are small in experimental realizations of the model; this slightly modifies the original Haldane model, but does not alter its topological properties.}\label{Fig_one}\end{figure}

We consider a two-dimensional (2D) honeycomb lattice realizing the two-band Haldane model \cite{Haldane1988,Jotzu2014}. The actual configuration used for our calculations, the so-called brickwall geometry, is depicted in Fig.~\ref{Fig_one}. Neglecting the effects of inter-particle interactions, and in the absence of any external trapping potential, the tight-binding Hamiltonian is given by 
\begin{align}
\hat H =& - t_{\text{NN}} \sum_{\langle j,k \rangle} \hat a^{\dagger}_j \hat a_k + t_{\text{NNN}} \sum_{\langle \langle j,k \rangle \rangle} i^{\circlearrowleft} \hat a^{\dagger}_j \hat a_k ,\label{Eq:Haldane}\\
&+(\DAB/2) \left ( \sum_{j \in \text{A sites}} \hat a^{\dagger}_j \hat a_j - \sum_{j \in \text{B sites}} \hat a^{\dagger}_j \hat a_j \right ), \notag
\end{align}
where $\hat a^{\dagger}_j$ creates a particle at lattice site $j$, $t_{\text{NN}}$ [resp.~$t_{\text{NNN}}$] is the tunneling amplitude for hopping processes between nearest-neighboring (NN) [resp.~next-nearest-neighboring (NNN)] sites, and $i^{\circlearrowleft}\!=\!\pm i$ depending on the orientation of the NNN hopping, e.g. $i^{\circlearrowleft}\!=\!+ i$ for clockwise hopping (see colored arrows in Fig.~\ref{Fig_one}). We emphasize that the chirality imposed by this orientation-dependent hopping term is responsible for the breaking of time-reversal symmetry (TRS) in the model, in analogy with the Lorentz force induced by an external magnetic field. Similarly to the traditional QH effect, this TRS-breaking term opens a gap in the two-band spectrum,  and favors bulk bands with non-trivial Chern numbers $C \!=\!\pm1$, see Ref.~\cite{Haldane1988} and below. In the second line of Eq.~\eqref{Eq:Haldane}, we introduced an offset $\DAB$ between $A$ and $B$ sites [Fig.~\ref{Fig_one}], which breaks inversion symmetry in the model: This term also opens a gap in the spectrum, but it favors bulk bands with trivial Chern numbers $C \!=\!0$. The competing effects associated with these two terms become evident when writing the Hamiltonian~\eqref{Eq:Haldane} in momentum representation,
\begin{align}
&\hat H(\bs k)= d_x (a \bs k) \hat \sigma_x + d_y (a \bs k) \hat \sigma_y + d_z (a \bs k) \hat \sigma_z, \\
&d_x (\bs k)=- t_{\text{NN}} \left ( \cos (k_x) + 2\cos k_y \right) ; \, d_y (\bs k)=- t_{\text{NN}} \sin k_x ;  \notag \\
&d_z (\bs k)= (\DAB/2) + 2 t_{\text{NNN}} \left [ \sin (k_x+k_y) - \sin (k_x-k_y) \right ].\notag
\end{align}
Here, we considered the brickwall geometry depicted in Fig.~\ref{Fig_one} and neglected purely-vertical NNN hopping, which is small in the experimental realisation of the Haldane model of Ref.~\cite{Jotzu2014}. In the absence of NNN hopping and offset ($t_{\text{NNN}}\!=\!\DAB\!=\!0$) the spectrum is gapless and displays conical intersections at the two Dirac points $\bs K_+=(0,\pi/3a)$ and $\bs K_-=(\pi/a,2\pi/3a)$. Including these two effects then potentially adds a mass term $M \hat \sigma_z$ to the effective Dirac equations associated with the two Dirac points $\bs K_\pm$, with masses given by $M_\pm\!=\!d_z (\bs K_\pm)$, respectively. On the one hand, the constant offset term $(\DAB/2) \hat \sigma_z$ generates the same mass term at the two Dirac points, with effective mass $M_\pm\!=\!(\DAB/2)$: This opens a topologically-trivial bulk gap in the spectrum, since the Chern numbers of the two bulk bands are given by $C\!=\!\pm(1/2) [\text{sign}(M_+) - \text{sign}(M_-)]\!=\!0$, see Ref.~\cite{Haldane1988}. On the other hand, adding the $\bs k$-dependent term associated with the chiral NNN-hopping generates opposite mass terms at the two Dirac points, with masses $M_{\pm}\!=\!\pm 2 \sqrt{3} t_{\text{NNN}}$: This opens a bulk gap and generates two bands with non-zero Chern numbers $C\!=\!\pm 1$.

\begin{figure}[h!]
\includegraphics[width=9.3cm]{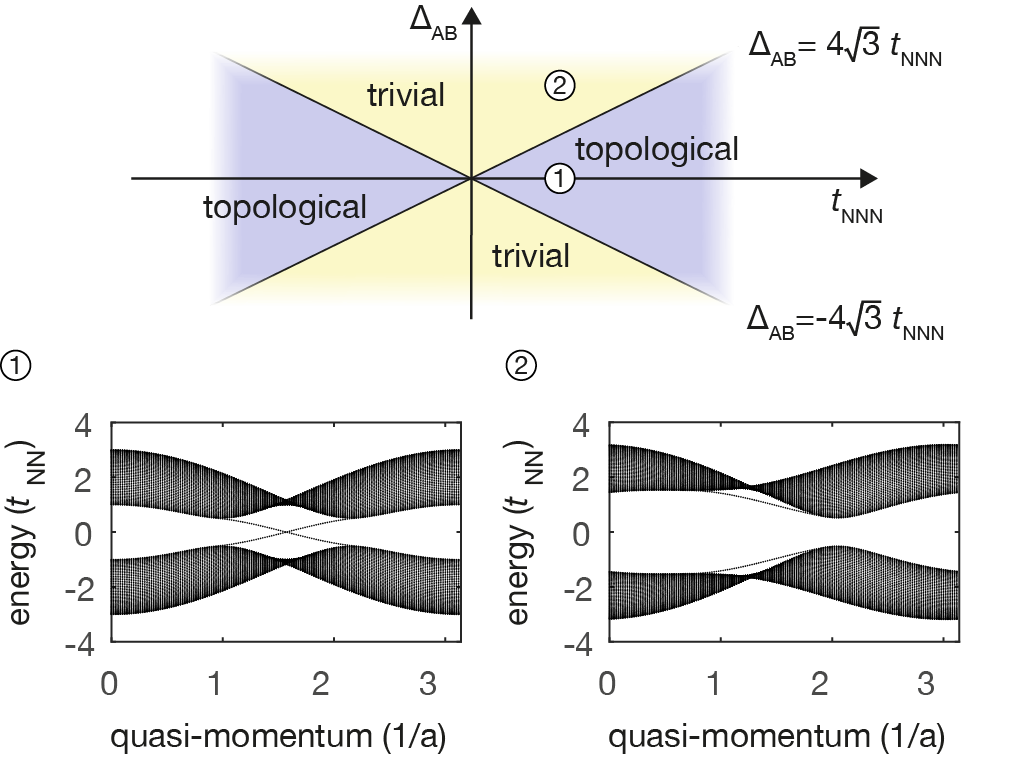}
\vspace{-0.cm} \caption{Topological phase diagram and two representative spectra. (top) The boundary lines in the phase diagram indicate gap-closing events in the two-band spectrum, through which the Chern number of the bulk bands is allowed to change. (bottom) Spectrum $E\!=\!E(k_y)$ of the Haldane model \eqref{Eq:Haldane}, set on a cylinder geometry aligned along $x$: \textcircled{1} $\DAB\!=\!0$ and  \textcircled{2} $\DAB\!=\!2 \Delta_{\text{trans}}$; here $t_{\text{NNN}}\!=\!0.15 t_{\text{NN}}$ and $\Delta_{\text{trans}}\!\equiv\! 4 \sqrt{3} t_{\text{NNN}}$. The situation in  \textcircled{1} corresponds to a topological band configuration, where the bulk bands are associated with Chern numbers $C \!=\!\pm1$, and where the bulk gap hosts a single edge-state mode (per edge); note that the two dispersion branches, with opposite group velocity (chirality), correspond to edge modes located on opposite edges of the cylinder. The case in  \textcircled{2} is a trivial band insulator with $C \!=\!0$. These two situations are indicated in the topological phase diagram.}\label{Fig_two}\end{figure}

The topology of the system is thus determined by the competition between these two opposite effects. For instance, keeping $t_{\text{NNN}}$ fixed, a variation of the offset $\DAB$ can be exploited to drive topological phase transitions: These are marked by a closing of the bulk gap ($M_{\pm}\!=\!0$) and a change in the Chern number of the bands. Noting that the mass terms at the two Dirac points are given by $M_{\pm}\!=\!(\DAB/2) \pm 2 \sqrt{3} t_{\text{NNN}}$, one finds that the critical offset at which a topological phase transition occurs is given by $\DAB\!=\!\pm \Delta_{\text{trans}}$, where $\Delta_{\text{trans}}\!\equiv\! 4 \sqrt{3} t_{\text{NNN}}$; see Fig.~\ref{Fig_two} for an illustration of the topological phase diagram associated with the model.

The manifestation of topology, and the related phase transitions, are directly visible when analyzing the model in Eq.~\eqref{Eq:Haldane} on a cylinder, e.g.~by applying periodic boundary conditions along the $y$ direction only. The corresponding spectrum $E\!=\!E(k_y)$, represented as a function of the quasi-momentum $k_y$, is shown in Fig.~\ref{Fig_two} for $\DAB\!=\!0$ and $\DAB\!=\!2 \Delta_{\text{trans}}$, keeping $t_{\text{NNN}}=0.15t_{\text{NN}}$ fixed. In the absence of offset, the bulk gap hosts two edge-state branches, namely, a single edge-state mode per edge of the cylinder. 
Importantly, the edge mode associated with a given edge has a well-defined chirality, $\text{sign}(\partial_{k_y}E)$, which describes the orientation of propagation along this edge  [Fig.~\ref{Fig_two}\textcircled{1}]. These edge modes are topologically protected, in that they cannot be removed by weak perturbations that preserve the bulk gap. 
Crossing the topological phase transition, i.e.~$\vert\DAB\vert\!>\!\vert\Delta_{\text{trans}}\vert$, the edge modes disappear from the bulk gap \footnote{
In Fig.~\ref{Fig_two}\textcircled{2}, some edge states, which are clearly distinguishable from the bulk states, are still visible  in the spectrum. However, as they are located away from the bulk gap, they are not topologically protected (these can be removed without changing the topology of the bulk bands). In addition, we find that their presence depends on the specific details of the boundary, e.g.~the orientation of the  boundaries with respect to the lattice.
} and the system reduces to a trivial band insulator [Fig.~\ref{Fig_two}\textcircled{2}]. 

It is worth pointing out that for traditional physical realizations of the model, i.e.~for a finite-size lattice defined on a 2D plane, the system only displays a single edge, and hence, a single topologically-protected edge mode (when parameters are set in the topological regime). We also refer the reader to Ref.~\cite{Lacki2016}, where a scheme to engineer cylindrical optical lattices has been proposed.

\subsection{Motivations behind the choice of the model}\label{section:motivations}

The topological interfaces and detection methods that we are about to discuss can be applied to a wide family of physical platforms featuring topological band structures. However, it is worth mentioning that the model considered in this work [Eq.~\eqref{Eq:Haldane}] does present several advantages. First of all, the Haldane model (and its variants) has been recently implemented in photonics and cold-atom experiments \cite{Rechtsman2013,Jotzu2014}, where the relevant model parameters ($t_{\text{NNN}}$, $\DAB$) can be finely tuned. Moreover, the minimal topological-band structure associated with the two-band Haldane model, which displays a single topologically-protected edge mode (per edge), significantly simplifies the analysis of edge-state physics: Indeed, a state that is prepared in the vicinity of an edge (or more generally, close to an interface separating topologically-different regions) will necessarily project unto two types of eigenstates: (1) bulk states, and (2) edge states that are associated with a well-defined chirality. This is in contrast with models displaying many bands, e.g.~the Hofstadter model \cite{Hofstadter1976}, where the different bulk gaps host edge-state modes of different chirality, which can all be potentially populated when preparing the initial state close to an edge/interface. We finally point out that the efficiency with which edge states are populated depends on various parameters, e.g.~the Fermi energy or the mean quasi-momentum of a prepared wave packet [the edge-mode dispersions shown in Fig.~\ref{Fig_two}\textcircled{1} are local in momentum space]. This latter aspect will be illustrated below.

\section{Creating topological interfaces}\label{sect:space_dep}

\subsection{The general strategy}\label{Section:general_strategy}

The topological properties of the system [Eq.~\eqref{Eq:Haldane}] have been discussed above for a homogeneous configuration of the parameters $t_{\text{NN}}$, $t_{\text{NNN}}$ and $\DAB$, and in the absence of any external trapping potential. In this case, the topological edge modes identified in Fig.~\ref{Fig_two}\textcircled{1} are located at the interface between the topological system [associated with non-zero Chern numbers $C \!=\!\pm1$] and vacuum [associated with trivial topology, $C \!=\!0$]: The edge states are  \emph{located at the edges of the system}. However, it is possible to engineer interfaces, separating different topologically-ordered regions, \emph{within the system}, as we now explain. 

%%%moved up 
%The general strategy can be simply described using a local-density-approximation (LDA) argument. Let us suppose that one parameter of the model is now varied continuously in space, e.g. $\DAB(\bs x)$, while keeping all other parameters fixed. Then, in an LDA approach, one can estimate the band structure locally, in a region located around some position $\bs x$, based on the value $\DAB(\bs x)$ that the spatially-varying parameter takes there. Following the topological phase diagram in Fig.~\ref{Fig_two}(a), one finds that different regions of the lattice can then be associated with different topological phases [i.e.~the bulk bands that are evaluated locally can have zero or non-zero Chern numbers depending on the value $\DAB(\bs x)$]. In particular, some singular regions are associated with a (local) gapless band structure: This occurs when $\DAB(\bs x)\!=\!\Delta_{\text{trans}}$, which defines the local topological interfaces \emph{within} the lattice. Since the band structure is locally gapless at these interfaces, these regions host topologically-protected modes, as will be investigated below.  In the following, we will loosely refer to these localized states as ``edge states", although we emphasize that these actually correspond to  modes that propagate within the lattice, i.e.~along a topological interface induced by the spatially-varying parameter $\DAB(\bs x)$. Similarly, the specific energy range associated with these localized modes will be referred to as the ``bulk gap", the ``bulk bands" being associated with states that are spatially delocalized.

A first proposal in Ref.~\cite{GoldmanPRL2010} suggested to achieve topological interfaces by locally (but strongly) modifying  hopping parameters in a central region of the system. This local change of hopping parameters, which can indeed split the system into topologically distinct regions, is particularly suitable for atom-chip implementations, where these parameters are set by tunable (and local) current-carrying wires; see also Refs.~\cite{Tenenbaum:2013,Reichl:2014}. 

\begin{figure}[h!]
\includegraphics[width=9.3cm]{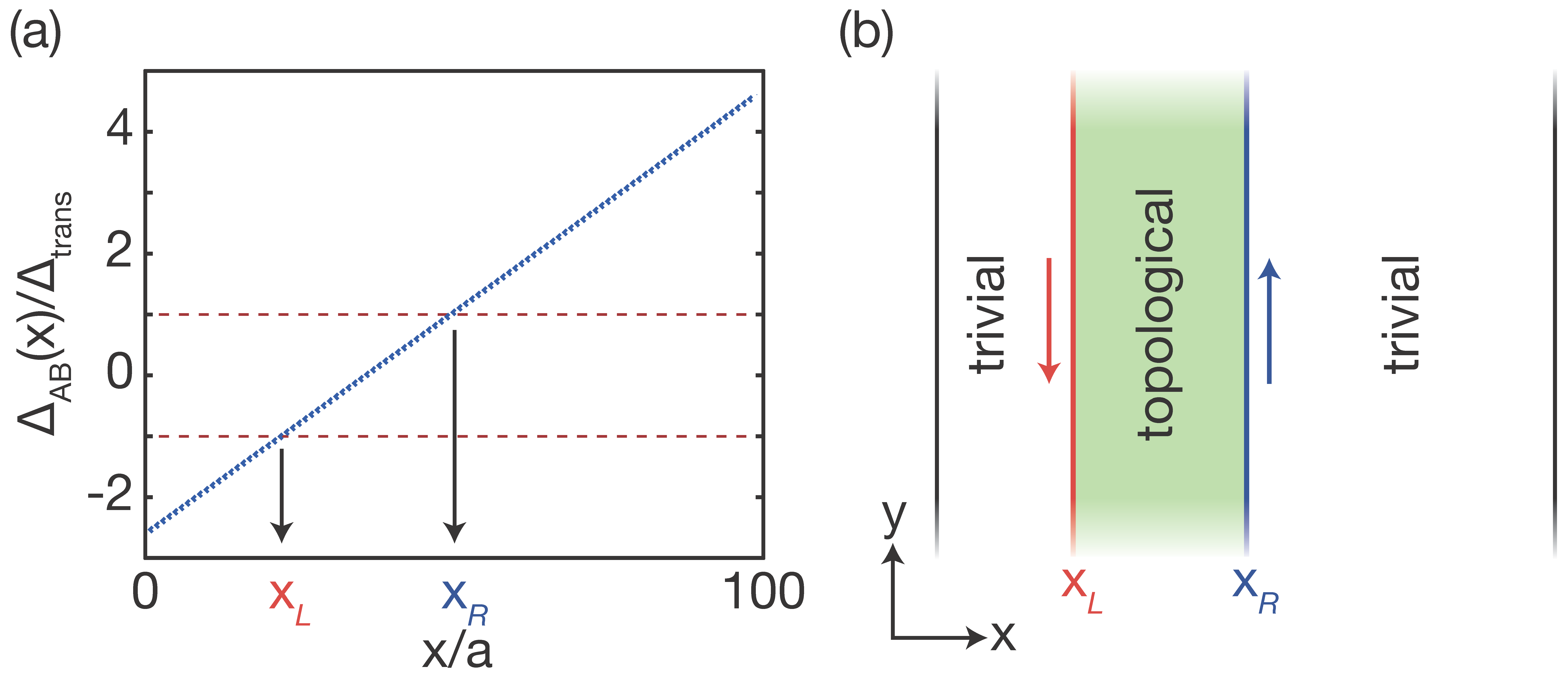}
\vspace{-0.cm} \caption{(a) Space-dependent offset $\DAB(x)$, as defined in Eq.~\eqref{space_offset}, for $\delta_{\text{max}}\!=\!20t_{\text{NN}}$, $t_{\text{NNN}}\!=\!0.4 t_{\text{NN}}$ and $L_x=100a$. The offset $\DAB(x)$ is expressed in units of the critical value $\Delta_{\text{trans}}\!\equiv\! 4 \sqrt{3} t_{\text{NNN}}$, at which topological transitions occur (see horizontal dotted lines). (b) The corresponding separation between topological and trivial regions, as dictated by the two interfaces located at $x\!=\!x_R\!=\!L_x/2$ and at $x\!=\!x_L$; see Eq.~\eqref{left_edge}. The inner region of the lattice corresponds to the topological region ($-\Delta_{\text{trans}}\!<\!\DAB(x)\!<\!\Delta_{\text{trans}}$), while the regions outside these frontiers are topologically trivial. Chiral topological edge modes are localized and propagate in the vicinity of these two interfaces, as indicated by the colored arrows. }\label{Fig_three}\end{figure}

In this work, we explore a different strategy, more practical for current optical-lattice experiments, which is based on the introduction of a spatially dependent offset $\DAB(\bs x)$ between neighboring sites. We take the offset to be a \emph{linear} function of one of the spatial coordinates, i.e.~we consider an offset of the form
\be
\DAB(x)=\left (\delta_{\text{max}}/L_x \right ) x + \Delta_{\text{trans}} - \left (\delta_{\text{max}}/2 \right ) , \label{space_offset}
\ee
where we introduced $L_x$, the system length along the $x$ direction, and the parameter $\delta_{\text{max}}$, which determines the slope of the space-dependent offset. 
In Section \ref{Section:experiment_lin}, we will show how such a spatial variation can be implemented experimentally, through a direct modification of the lattice potential.
Note that the function $\DAB(x)$ in Eq.~\eqref{space_offset} is chosen such that the critical value,
\be
 \Delta_{\text{trans}}\!=\!\DAB(x\!=\!L_x/2),
 \ee 
 is exactly reached \emph{at the center of system}, i.e.~where the external trapping potential $V_{\text{trap}}(\bs x)$ is minimal [Sec.~\ref{section_trap}]. This critical position will be denoted $x_R\!=\!L_x/2$, as it defines the location of the \emph{right interface}, where topologically-protected chiral modes are localized. 
% Indeed, since the system locally reaches a topological transition point (i.e.~a local gap-closing event) at $x\!=\!x_R$, we expect the existence of well-localized states in the vicinity of this region $x\!\approx\! x_R$. Moreover, since the Chern number of the lowest bulk band locally changes throughout the transition point ($C=\pm 1 \leftrightarrow 0$), we expect these localized states to be chiral, as in the standard case of a Chern insulator surrounded by vacuum [Fig.~\ref{Fig_two}(b)]. Summarizing, the interface defined at $x\!=\!x_R$ hosts localized states with a well-defined orientation of propagation, as illustrated in Figs.~\ref{Fig_three}(a)-(b).
Note that the other critical value $-\Delta_{\text{trans}}$ can also be reached within the system, at the position
\be
x_L=L_x \left ( \frac{1}{2} - \frac{2 \Delta_{\text{trans}}}{\delta_{\text{max}}}   \right )  =  x_R - \left (\frac{8 \sqrt{3} L_x }{\delta_{\text{max}}} \right ) t_{\text{NNN}},  \label{left_edge}
\ee
which defines the location of the \emph{left interface}; see Figs.~\ref{Fig_three}(a)-(b). 

At this stage, it is important to note that we are dealing with a competition between two relevant effects. On the one hand, increasing the ratio $ t_{\text{NNN}}/\delta_{\text{max}}$ allows to spatially separate the edge modes located on different interfaces [Eq. \eqref{left_edge}], which is an important feature in order to detect clean chiral edge-state propagation (and potentially, to limit back-scattering processes in the presence of engineered disorder). On the other hand, increasing the slope $\delta_{\text{max}}$ allows to improve the localization of the edge states within each interface, as shown below. 

\begin{figure}[h!]
\includegraphics[width=9.3cm]{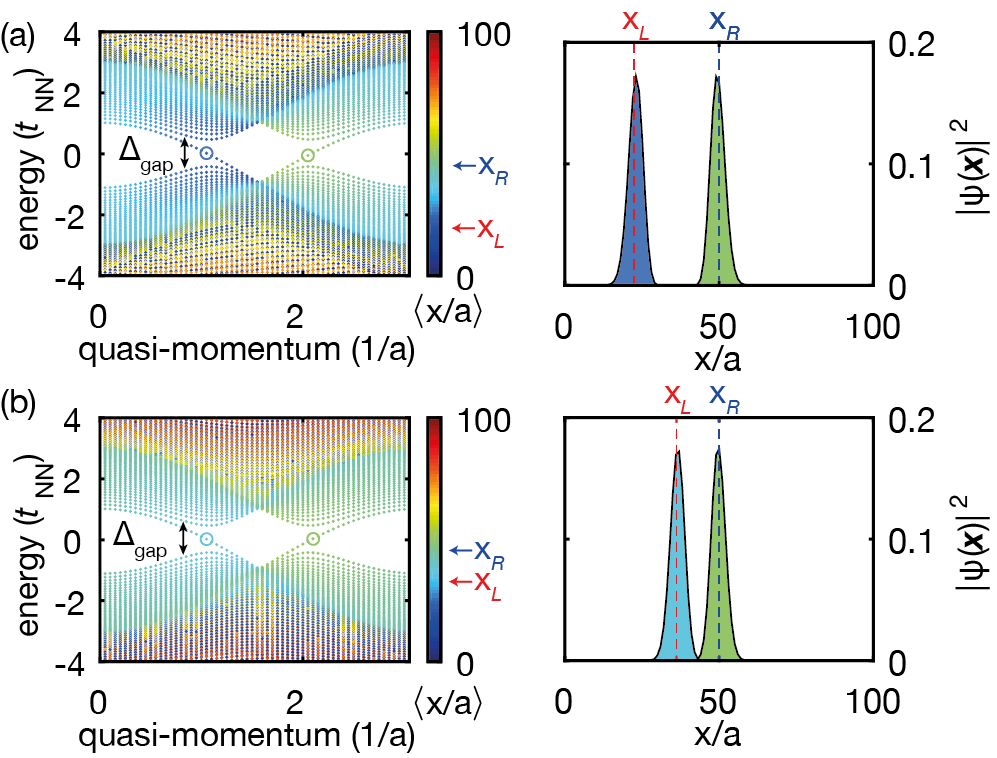}
\vspace{-0.cm} \caption{Energy spectrum $E\!=\!E(k_y)$, as a function of quasi-momentum $k_y$, and edge-state amplitudes for two values of the NNN hopping amplitude: (a) $t_{\text{NNN}}\!=\!0.4 t_{\text{NN}}$, (b) $t_{\text{NNN}}\!=\!0.2 t_{\text{NN}}$. The system is solved on a cylinder of length $L_x\!=\!100 a$, and with a space-dependent offset $\DAB(x)$, Eq.~\eqref{space_offset}, characterized by $\delta_{\text{max}}\!=\!20 t_{\text{NN}}$. Each eigenvalue $E(k_y)$ is colored in terms of the mean position $\langle x \rangle$ of its corresponding eigenstate (see colorbars). The eigenstates represented on the right correspond to the eigenenergies $E(k_y)$ indicated by blue and green circles in the spectrum, respectively. Vertical dotted lines in the right panels correspond to the interface positions, $x_{L,R}$, as predicted by Eqs.~\eqref{space_offset}-\eqref{left_edge}. We note that the group velocity $v_g^y$ associated with the edge-modes remains approximately constant as $t_{\text{NNN}}$ is increased. Moreover, the bulk bands are only slightly distorted in the vicinity of the bulk gap, whose size is approximatively $\Delta_{\text{gap}}\!\approx\!0.9 t_{\text{NN}}$. We point out that only the states located within the bulk gap, with an approximately linear dispersion relation, are well (spatially) localized. }\label{Fig_4}\end{figure}

\begin{figure}[h!]
\includegraphics[width=9.3cm]{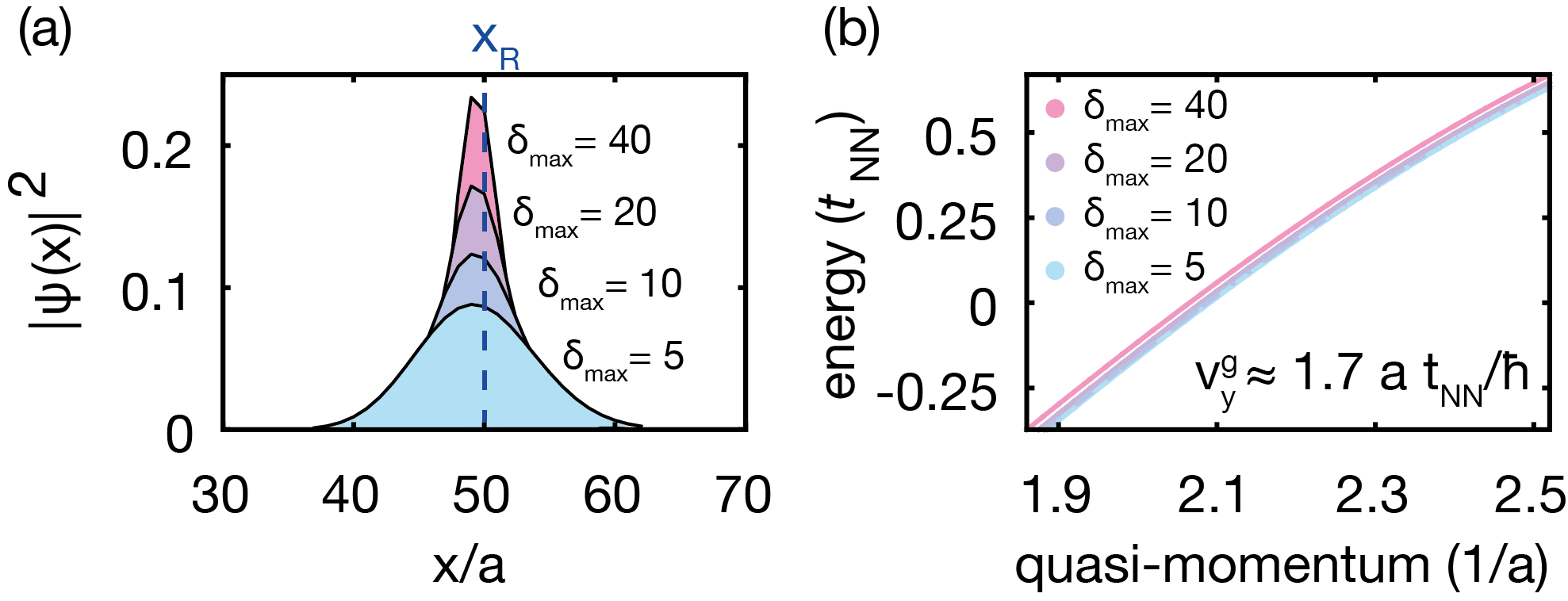}
\vspace{-0.cm} \caption{(a) Amplitude of an edge state localized around $x_R$ for $t_{\text{NNN}}\!=\!0.4 t_{\text{NN}}$, and four different values of the slope parameter $\delta_{\text{max}}$, as indicated on the figure. The edge mode propagating along the central interface is more localized as $\delta_{\text{max}}$ is increased. (b) Dispersion relation $E\!=\!E(k_y)$ of the edge mode at $x_R$, for the same values of the system parameters. The group velocity of this localized mode is found to be stable around the value $ v_g^y \approx 1.7 a t_{\text{NN}}/\hbar$. Here the system size along the $x$ direction is $L_x\!=\!100 a$. }\label{Fig_5}\end{figure}

We demonstrate these two competing effects by diagonalizing the system on a cylinder aligned along the $x$ direction, noting that the system still preserves translational symmetry along the $y$ direction.  The corresponding spectrum, as well as the amplitude $\vert \psi (x) \vert^2$ of two representative states, are represented in Fig.~\ref{Fig_4} for two different values of the parameter $ t_{\text{NNN}}$. First of all, we note that the states represented in Fig.~\ref{Fig_4}(a) are well localized in the vicinity of the interfaces located at $x\!=\!x_{L,R}$ and that the dispersion relation of these modes, $E\!=\!E(k_y)$ are reminiscent of those associated with standard topological edge states [Fig.~\eqref{Fig_two}\textcircled{1}]: These dispersions are approximately linear and they are well isolated from the bulk bands associated with delocalized states (the size of the corresponding ``bulk gap" is found to be $\Delta_{\text{gap}}\!\approx\!0.9 t_{\text{NN}}$ in Fig.~\ref{Fig_4}). Hence, these dispersions describe one-dimensional Dirac fermions, propagating along the $y$ direction, with an approximately constant group velocity $+v_g^y$ [resp.~$-v_g^y$], along the right [resp.~left] interface. Besides, we note that these edge states have a localization length of about five lattice sites for the (realistic) parameters chosen in these calculations [$\delta_{\text{max}}\!=\!20t_{\text{NN}}$ and $L_x=100a$]. Figure~\ref{Fig_4}(b) shows that reducing the parameter $ t_{\text{NNN}}$ affects the spatial separation between the two localized modes, as predicted by Eq.~\eqref{left_edge}; however, quite surprisingly, we observe that this change only very slightly modifies the localization length and dispersion (group velocity) of the modes. Importantly, the localization length of the localized mode propagating along the central interface $x\!=\!x_R$ is significantly reduced by increasing the slope parameter $\delta_{\text{max}}$, as we illustrate in Fig.~\ref{Fig_5}(a). We further show in Fig.~\ref{Fig_5}(b) the robustness of the group velocity  $v_g^y$ against changes in the slope parameter $\delta_{\text{max}}$, which stabilizes around the value $ v_g^y \approx 1.7 a t_{\text{NN}}/\hbar$.  

\begin{figure}[h!]
\includegraphics[width=9.3cm]{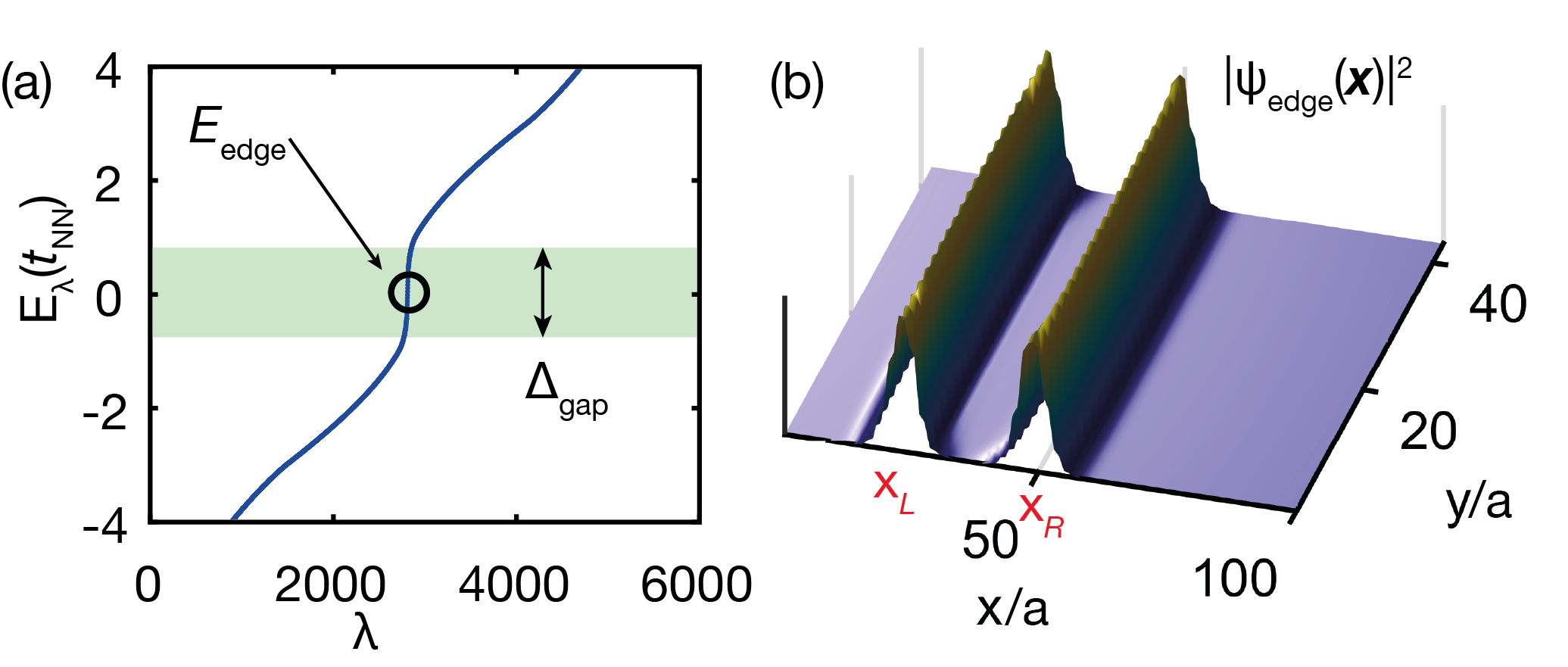}
\vspace{-0.cm} \caption{(a) Energy spectrum for a finite-size open-boundary honeycomb lattice [Eq.~\eqref{Eq:Haldane}], with spatially-dependent offset given by Eq.~\eqref{space_offset}. Here $\lambda$ is an integer labeling the eigenvalues by increasing order.  The bulk gap has a size $\Delta_{\text{gap}}\!\approx\! t_{\text{NN}}$, as indicated by the green region where the density of states is strongly reduced. (b) Amplitude of a representative eigenstate $\vert \psi_{\text{edge}} (\bs x)\vert^2$, with energy $E_{\text{edge}}\!\approx\!0$ in the bulk gap. This state is localized on the topological interfaces located at $x\!=\!x_{R,L}$, with a localization length of about five lattice sites [in agreement with Fig.~\ref{Fig_4}(a)]. System parameters are $t_{\text{NNN}}\!=\!0.4 t_{\text{NN}}$, $\delta_{\text{max}}\!=\!20 t_{\text{NN}}$, $L_x\!=\!100a$ and $L_y\!=\!56a$. }\label{Fig_2D_states}\end{figure}

Finally, we diagonalized the full 2D open-boundary lattice, and we present the corresponding spectrum and a representative eigenstate in Fig.~\ref{Fig_2D_states}.  The spectrum is shown in Fig.~\ref{Fig_2D_states}(a), where the presence of the bulk gap is identified through a severe reduction of the density of states around $E\!\approx\!0$. Note that this spectrum is in agreement with the one presented in Fig.~\ref{Fig_4}, for the cylinder-geometry case. Fig.~\ref{Fig_2D_states}(b) shows a representative eigenstate, whose energy $E_{\text{edge}}\!\approx\!0$ is located within the bulk gap. We find that the states present in the bulk gap are indeed well localized on the topological interfaces located at $x\!=\!x_{R,L}$. 

This method of displacing the conducting topological interfaces by tuning $t_{\text{NNN}}$ raises an interesting possibility for technological applications: 
In the realization of the Haldane model based on Floquet engineering, the value of $t_{\text{NNN}}$ is controlled by the frequency, amplitude and polarization of an oscillating external force \cite{Oka2009,Jotzu2014}. 
Indeed, the original proposal considered circularly polarized light illuminating a sheet of graphene.
If such a sheet were to be placed on a substrate with an incommensurate lattice spacing (see Sec.~\ref{Section:experiment_lin} and Refs.~\cite{Fain1980,Tang2013,Woods2014}), this may lead to a spatially varying site-offset.
Then, tuning the properties of the illumination could be used as a method for dynamically displacing conducting channels in a material.

%In an experiment, the typical size of an atomic cloud is about 80 sites diameter. In order to obtain a detectable edge-state signal, an edge-state localization length of about 5-20 lattice sites is preferable. We find that this can be achieved, together with a clear separation between the interfaces (about 100 sites), by choosing the system parameters values in the realistic ranges $t_{\text{NNN}}\!=\!0.25\!-\!0.5 t_{\text{NN}}$ and $\delta_{\text{max}}\!=\!10\!-\!20 t_{\text{NN}}$. As discussed above, the edge-state velocity is found to be almost unaffected by the choice of these parameters, and we find $\vert v_g \vert \approx 1.73 a t_{\text{NN}}/\hbar$. 

\subsection{On the effects of time-reversal-symmetry breaking and spatially-resolved edge states}\label{Section:TRS}

In the previous section, we demonstrated that increasing the NNN hopping parameter allows one to spatially separate the topological interfaces induced by $\DAB(x)$, and hence, resolve the edge modes propagating with opposite chirality in the system. 
We now address a natural question: ``{\emph{What happens when $ t_{\text{NNN}}\!=\!0$, namely when time-reversal-symmetry is present in the system?}". 
In this situation, we note that $\DAB(x)\!>\!0$ in the region defined by $x\!>\!L_x/2$: The system is locally gapped, with positive mass terms at both Dirac points (the gap is topologically trivial, as it is only due to local inversion-symmetry breaking). 
In the other region,  $x\!<\!L_x/2$, the system is also locally gapped since $\DAB(x)\!<\!0$, but now with negative mass terms at both Dirac points. 
We deduce from this that the full system is topologically trivial, but that the mass terms are locally reversed at the center of the system $x\!=\!x_R$. 
Consequently the system is locally gapless at $x\!=\!x_R$, where $\DAB(x_R)\!=\!t_{\text{NNN}}\!=\!0$; see also the phase diagram in Fig.~\ref{Fig_two}. 
Hence, when setting $t_{\text{NNN}}\!=\!0$, there is a single ``trivial interface" at $x\!=\!x_R$, where states are localized (due to the fact that, in the LDA picture, the system is gapless there); see the sketch in Fig.~\ref{Fig_TRS}(a) and the numerical calculation (spectrum and localized states) shown in Fig.~\ref{Fig_TRS}(b). 
This interface, however, is of a different nature than the ones discussed above for $t_{\text{NNN}}\!\ne\!0$, since time-reversal-symmetry is satisfied when $t_{\text{NNN}}\!=\!0$: 
The single interface now hosts localized edge modes with both chirality (i.e.~the interface is equivalent to a standard, isolated, 1D lattice) meaning that conduction in these modes is not protected from back-scattering. 
In Ref. \cite{Leder2016}, a related type of localized state was created in a \emph{one-dimensional} lattice, by locally reversing the mass of a 1D Dirac point.

Note that the pair of edge modes shown in Fig.~\ref{Fig_4} and~\ref{Fig_TRS}(b) can be resolved in $k$-space for all values of $t_{\text{NNN}}$, including for the time-reversal-invariant case $t_{\text{NNN}}\!=\!0$. This indicates that edge modes with opposite chirality can still be individually selected on this interface, even in the time-reversal-invariant case, e.g.~by tuning the mean quasi-momentum of a Gaussian wave packet prepared at the center of the system. For instance, in the TRS case represented in Fig.~\ref{Fig_TRS}(b), tuning the mean quasi-momentum to the value $k_y^0\!\approx\!+2.1/a$ will mostly project the wave packet unto the localized states with positive group velocity $v_g^y\!>\!0$. We demonstrate this selectivity of localized chiral edge modes, for the singular TRS case, in Appendix \ref{Appendix:TRS}, where the dynamics of Gaussian wave packets are obtained through numerical simulations of the full 2D lattice.

This analysis has an important corollary:~Showing the uni-directional propagation of states along the central interface is not enough to demonstrate the existence of topologically-non-trivial edge modes in the system. Indeed, it is important to prove that a non-trivial (spatially separated) interface only hosts modes with a given chirality. A possible protocol for verifying this consists in initially preparing a wave packet on an interface (e.g.~around $x\!=\!x_R$), and imaging the time-evolution of the wave packet for various values of the mean quasi-momentum (i.e.~scanning the full Brillouin zone): In the non-trivial-topological situation, there should only be a single interval of values $k_y^0$ that gives rise to an unidirectional motion along this specific interface. This would unambiguously demonstrate the chirality associated with this interface. Then one could perform the same analysis on the other interface ($x_L$), and demonstrate that the observed motion has the opposite chirality in that case. In photonic and mechanical systems, adding dislocations to topological edge states has been used as a method to probe their chirality \cite{Rechtsman2013,Susstrunk:2015}.

 \begin{figure}[h!]
\includegraphics[width=9.cm]{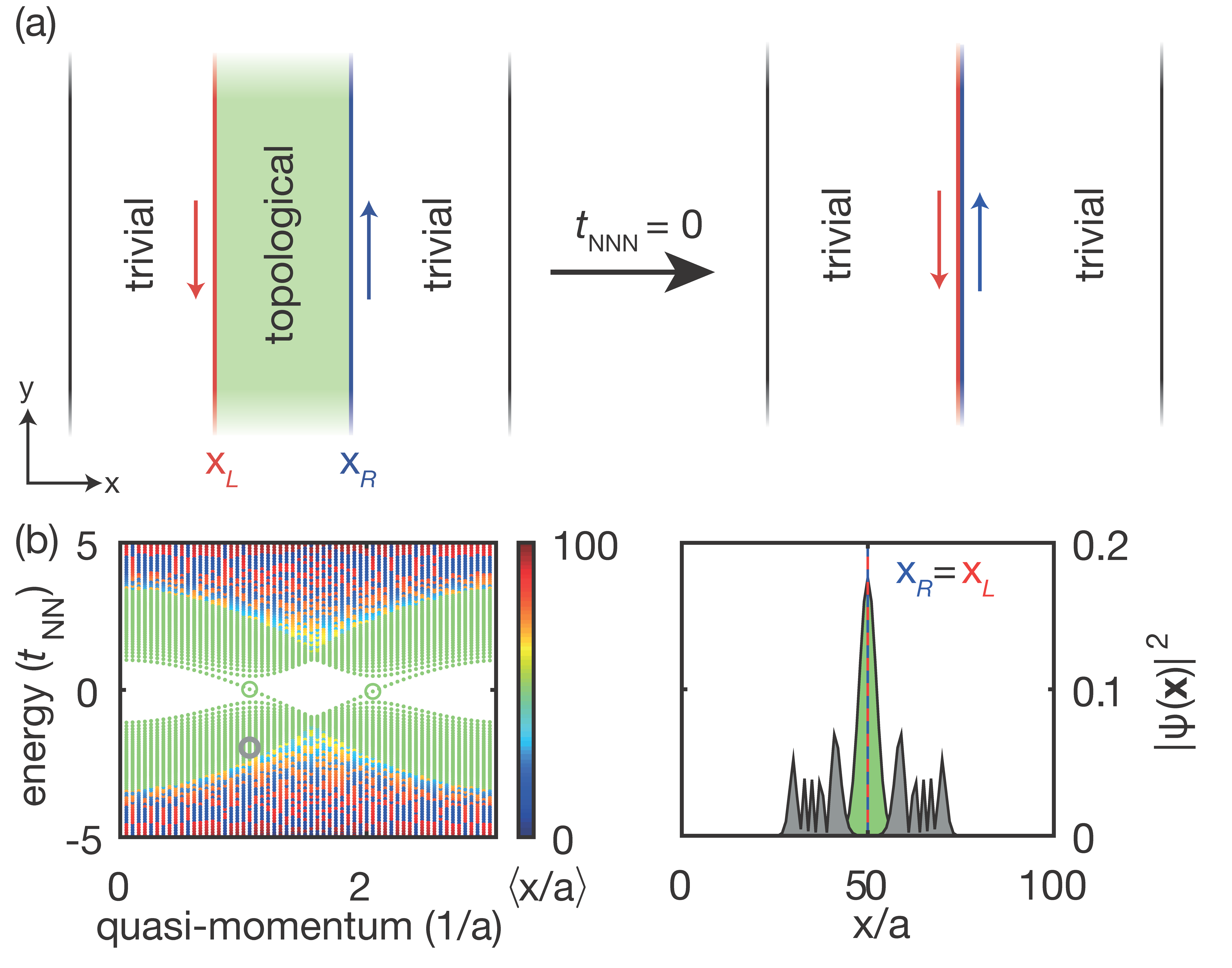}
\vspace{-0.cm} \caption{(a) In the absence of NNN hopping, $t_{\text{NNN}}\!=\!0$, the system is invariant under time reversal: The topological region disappears, which is indicated here by the equality $x_R\!=\!x_L$ for $t_{\text{NNN}}=0$; see Eq.~\eqref{left_edge}. The trivial ``interface" at $x\!=\!L_x/2$ hosts localized propagating states, with opposite chirality: In this scheme, the chirality of localized modes can only be spatially resolved when setting $t_{\text{NNN}}\!\ne\!0$, i.e.~when time-reversal symmetry is broken [see Figs.~\ref{Fig_three} and~\ref{Fig_4}]. (b) Energy spectrum $E\!=\!E(k_y)$, as a function of quasi-momentum $k_y$, and amplitude of two localized states (green) and one delocalized state (grey) for the time-reversal-invariant case $t_{\text{NNN}}\!=\!0$. The energy of these depicted states is indicated in the spectrum by two green circles and one grey circle, respectively. All other system parameters are the same as in Fig.~\ref{Fig_4}. We point out that only the states located within the bulk gap are well (spatially) localized [see the bulk state depicted in grey versus the localized states in green].  }\label{Fig_TRS}\end{figure}

\subsection{Adding the external harmonic trap}\label{section_trap}

In this section, we discuss the effects of a harmonic trapping potential $V_{\text{trap}} (\bs x)$ added on top of the lattice system introduced in Sec.~\ref{Section:general_strategy}. Since this is a typical feature of optical-lattice experiments, such a harmonic trap will be included in the numerical simulations presented in Sec.~\ref{Section:dynamics}-\ref{Section:disorderb}. The aim of this section is to identify realistic configurations of the trap for which the propagation of chiral edge modes can still be observed in experiments.

The robustness of topological edge states against smoothly varying potentials (e.g.~harmonic traps) has already been investigated numerically in Refs.~\cite{StanescuPRA,GoldmanPRL,GoldmanEPJST,GoldmanPNAS,CocksPRA}. These results can also be understood based on a general LDA argument, which can be summarized as follows: A topological edge mode, spatially localized in some region $R$ in the absence of the external trap, will survive (in this region) in the presence of the trap $V_{\text{trap}} (\bs x)$ as long as \be
V_{\text{trap}} (\bs x\! \in R) \!<\! \Delta_{\text{gap}},
\ee 
where $\Delta_{\text{gap}}$ denotes the size of the bulk gap that hosts and protects the topological edge mode. This criterium signifies that the topological gap, locally estimated in the region of interest $R$, should not collapse due to the presence of the trap, in order to probe the propagation of the edge mode in this region.

One can directly apply this argument to the system introduced in Section~\ref{Section:general_strategy}, where the regions of interest correspond to the topological interfaces at $x\!=\!x_{L,R}$. To analyze this situation, let us write the harmonic potential in the following form
\be
V_{\text{trap}} (\bs x)= V_x \left ( \frac{x-x_R}{x_L\!-\!x_R} \right)^2 + V_y \left ( \frac{y-y_0}{L_{\text{obs}}} \right)^2 , \label{xy_trap}
\ee
where we introduced an ``observation" length $L_{\text{obs}}$, and where $(x_R,y_0)\!=\!(L_x/2,L_y/2)$ is the center of the trap. First of all, let us focus on the central interface located at $x\!=\!x_R$. This region effectively feels a one-dimensional harmonic trap, aligned along the propagation direction ($y$), of the form $V_{\text{trap}} (y)\!=\!V_y [(y-y_0)/L_{\text{obs}}]^2$. Hence, following the LDA argument above, we find that the detection of chiral modes, propagating along this interface over a distance $L_{\text{obs}}$,  is possible as long as $V_y\!<\!\Delta_{\text{gap}}$. A similar criterion can also be introduced if one is interested in the detection of the other topological interface [$x\!=\!x_L$], which sets a condition on the other trap parameter $V_x\!<\!\Delta_{\text{gap}}$. We verified these LDA predictions through a direct numerical diagonalization of the lattice Hamiltonian [Eq.~\eqref{Eq:Haldane},\eqref{space_offset}], in the presence of the trap [Eq.~\eqref{xy_trap}]. Figure \ref{Fig_harmonic} compares the shape of the harmonic potential $V_{\text{trap}} (\bs x)$ in the 2D plane with the amplitude $\vert \psi_{\text{edge}}(\bs x)\vert^2$ of a representative eigenstate, whose energy $E\!\approx\!0$ is located within the bulk gap. This figure shows that this ``edge" state is indeed well localized along the two topological interfaces, but only within regions where $V_{\text{trap}} (\bs x)\!<\!\Delta_{\text{gap}}\!\approx\!0.9 t_{\text{NN}}$; see the regions encircled by the white dotted ellipses in the right panels of Fig.~\ref{Fig_harmonic}. 

This analysis, which confirms the general LDA prediction, identifies the trapping-potential configurations for which clear chiral motion can be observed in an experimental realization of our topological-interface model [Eq.~\eqref{Eq:Haldane},\eqref{space_offset}]. In particular, it highlights the robustness of the chiral modes propagating \emph{at the center} of the trap, i.e.~along the engineered interface at $x\!=\!x_R$.

\begin{figure}[h!]
\includegraphics[width=9.3cm]{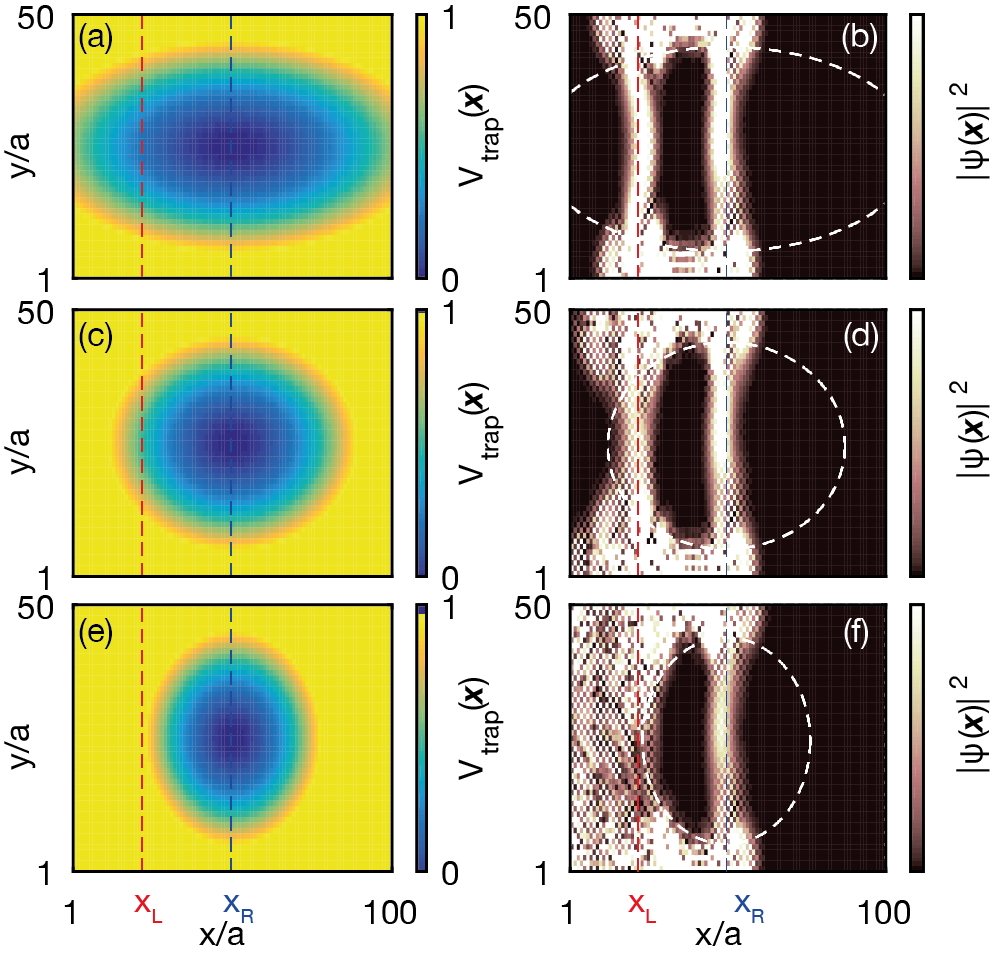}
\vspace{-0.cm} \caption{Comparison between the shape of the harmonic potential $V_{\text{trap}} (\bs x)$ in the 2D plane (left) and the amplitude $\vert \psi_{\text{edge}}(\bs x)\vert^2$ of an eigenstate whose energy is located within the bulk gap (right). The trap parameters are given by (a)-(b) $V_{x}\!=\!V_{y}\!=\!0.2 t_{\text{NN}}$, (c)-(d) $V_{x}\!=\!V_{y}\!=\!0.5 t_{\text{NN}}$, and (e)-(f) $V_{x}\!=\!V_{y}\!=\!1.0 t_{\text{NN}}$. In both cases, $L_{\text{obs}}\!=\!20a \sqrt{V_y/t_{\text{NN}}}$, so that the available propagation distance along the central interface is of about $40$ sites. The ``edge" state is indeed well localized along the topological interfaces, but only within regions where $V_{\text{trap}} (\bs x)\!<\!\Delta_{\text{gap}}\!\approx\!0.9 t_{\text{NN}}$, see the white dotted ellipses on the right panels. System parameters are $t_{\text{NNN}}\!=\!0.4 t_{\text{NN}}$, $\delta_{\text{max}}\!=\!20 t_{\text{NN}}$, $L_x\!=\!100a$ and $L_y\!=\!50a$. On the left panels, the trap $V_{\text{trap}} (\bs x)$ is measured in units of $t_{\text{NN}}$.}\label{Fig_harmonic}\end{figure}

\section{The radial-symmetric topological interface}\label{section:radial_interface}

In this section, we discuss how the space-dependent offset $\DAB(x)$ in Eq.~\eqref{space_offset} can be modified so as to generate a single radial-symmetric topological interface. As will be shown below, this configuration allows one to further reduce the effects of the external harmonic trap, and hence, to probe the physics of topologically-protected modes on potentially larger length scales (and longer time scales). 

In order to achieve such a radial-symmetric interface, we now consider that the space-dependent offset $\DAB(\bs x)$ can be created in the form of a radial-symmetric Gaussian
\begin{align}
\DAB(r) &= \frac{5 \Delta_{\text{trans}}}{2}  \left [ 1 - \exp ( - r^2 / 2 R_{\text{inter}}^2  ) \right ] ,\label{space_offset_radial}\\
 r &= \sqrt{(x-L_x/2)^2+(y-L_y/2)^2}, \nonumber
\end{align}
where we introduced the radius of the interface $R_{\text{inter}}$, which separates the (inner) topologically-non-trivial region from the (outer) trivial region; see also Ref.~\cite{Reichl:2014}. Indeed, the function $\DAB(r)$ in Eq.~\eqref{space_offset_radial} is chosen such that $\DAB\!=\!0$ at the center of the trap ($r\!=\!0$), then increases as a Gaussian function, and reaches the topological transition point in the vicinity of the radius $R_{\text{inter}}$, i.e.~$\DAB(r\!=\!R_{\text{inter}})\!\approx\!\Delta_{\text{trans}}$. Note that the location of this radial topological interface corresponds to the width of the Gaussian-shaped offset function $\DAB(r)$: in the vicinity of the critical radius $r\!\approx\!R_{\text{inter}}$, the offset $\DAB(r)$ depends approximately linearly on $r$, with a slope given by  
\be
\DAB'(r\!=\!R_{\text{inter}})\!\approx\!(3/2)\Delta_{\text{trans}}/R_{\text{inter}}.
\ee
Setting the radius parameter to be $R_{\text{inter}}\!\approx\! 20a$, one approximately recovers the slope associated with the linear-interface scheme illustrated in Fig.~\ref{Fig_three}(a). The experimental implementation of such an offset is discussed in Section~\ref{Section:experiment_circ}.

In this radial-symmetric configuration, the LDA argument predicts that the chiral edge mode is now localized along the radius $r\!=\!R_{\text{inter}}$, where it performs a circular motion. One estimates the angular velocity of this chiral mode to be given by $\dot \theta\!\approx\!v_g^y/R_{\text{inter}}$, where $v_g^y$ is the group velocity evaluated above for the linear-interface case; since the microscopic details of the boundary are changed, the velocity $v_g^y$ can potentially slightly differ from the linear case. 

We verified these predictions by performing a direct diagonalization of the full 2D lattice  described by Eqs.~\eqref{Eq:Haldane} and \eqref{space_offset_radial}. The corresponding energy spectrum, shown in Fig.~\ref{Fig_2D_states_radial}(a), indicates that the new shape of the interface does not significantly modify the bulk bands obtained in the linear-interface case:~In particular, the edge modes are still protected by a bulk gap of size $\Delta_{\text{gap}}\!\approx\! t_{\text{NN}}$. A representative edge state, whose energy is located within the bulk gap, is shown in Fig.~\ref{Fig_2D_states_radial}(b). We find that these edge states are indeed well localized around the radius $r\!=\!R_{\text{inter}}$, with a typical localization length of  about five lattice sites [similarly to the linear case]. We point out that, when comparing the radial and linear configurations in Fig.~\ref{Fig_2D_states_radial}(a), we set the corresponding system parameters in such a way that the offset slopes  are of the same order in both configurations, i.e.~$\delta_{\text{max}}/L_x\!\approx\!(3/2)\left (\Delta_{\text{trans}}/R_{\text{inter}} \right )$. 

\begin{figure}[h!]
\includegraphics[width=9.3cm]{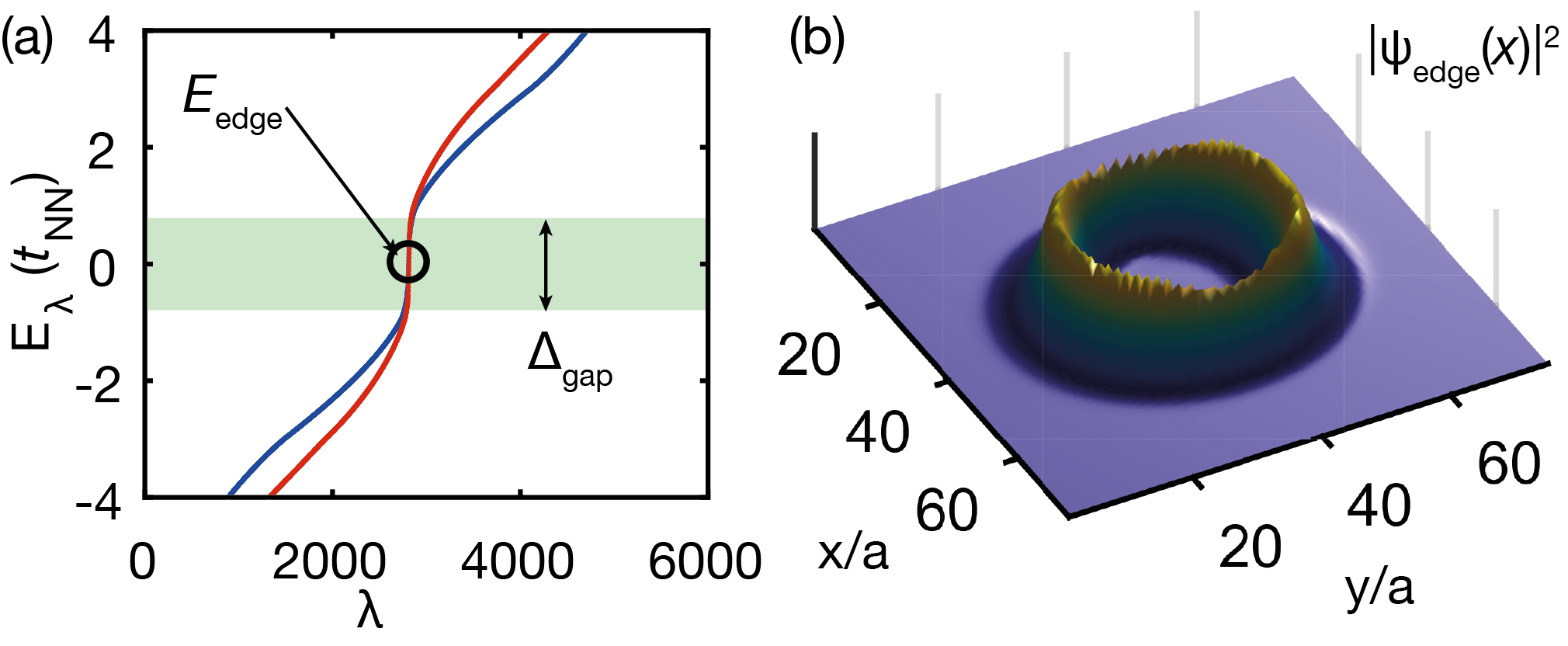}
\vspace{-0.cm} \caption{(a) Energy spectrum for the radial-interface (red) configuration [Eq.~\eqref{space_offset_radial}]; this spectrum is compared with the linear-interface case (blue). Here $\lambda$ is an integer labeling the eigenvalues by increasing order.  In both cases, the bulk gap has a size $\Delta_{\text{gap}}\!\approx\! t_{\text{NN}}$, as indicated by the green region where the density of states is strongly reduced. (b) Amplitude of a representative eigenstate $\vert \psi_{\text{edge}} (\bs x)\vert^2$, with energy $E_{\text{edge}}\!\approx\!0$ within the bulk gap, for the radial-interface configuration. This state is localized around the radial topological interface, defined by the radius $r\!=\!R_{\text{inter}}$. System parameters are $t_{\text{NNN}}\!=\!0.4 t_{\text{NN}}$, $R_{\text{inter}}\!=\!17 a$, and $L_x\!=\!L_y\!=\!75a$.}\label{Fig_2D_states_radial}\end{figure}

Finally, we discuss the effects of the harmonic trap in the radial-interface configuration. Following the discussion of Section~\ref{section_trap}, we write the trapping potential in the form
\be
V_{\text{trap}}(r)=V_0 (r/R_{\text{inter}})^2 . \label{trap}
\ee
In this case, the LDA argument predicts that the edge mode propagating along the radial interface $r\!=\!R_{\text{inter}}$ will survive in the presence of the trap, as long as $V_0\!<\!\Delta_{\text{gap}}$. We verified this statement by diagonalizing the lattice system in the presence of the trap, for various values of the trap parameter $V_0$. As an illustration, we present a perfectly conserved edge state in Fig.~\ref{Fig_harmonic_radial}, obtained for the parameters $V_0\!=\!0.25 t_{\text{NN}}$ and $R_{\text{inter}}\!=\!17a$. The remarkable robustness of this radial-symmetric edge mode relies on that it is \emph{entirely} located in a region where $V_{\text{trap}}(\bs x)\!<\!\Delta_{\text{gap}}$; see the white dotted ellipse in Fig.~\ref{Fig_harmonic_radial}(b). This constitutes a significant advantage, as compared to the linear-interface configuration [Fig.~\ref{Fig_harmonic}].

\begin{figure}[h!]
\includegraphics[width=11.cm]{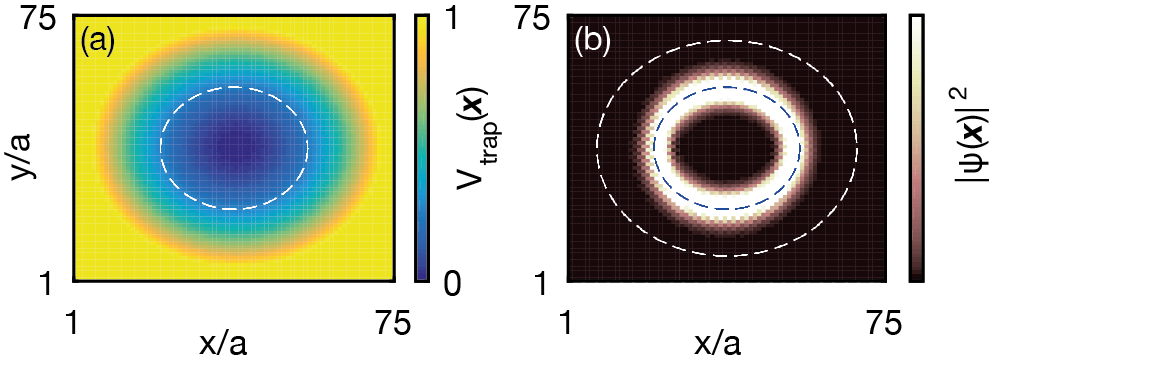}
\vspace{-0.cm} \caption{Comparison between (a) the shape of the harmonic potential $V_{\text{trap}} (\bs x)$ in the 2D plane, and (b) the amplitude $\vert \psi_{\text{edge}}(\bs x)\vert^2$ of an eigenstate whose energy is located within the bulk gap, for the radial-interface configuration [Eq.~\eqref{space_offset_radial}]. The radius of the circular interface $R_{\text{inter}}$ is indicated by a white [resp. blue] dotted circle in (a) [resp.~(b)]. The robustness of this localized edge state relies on that it is entirely located in a region where $V_{\text{trap}}(\bs x)\!<\!\Delta_{\text{gap}}$, see the white dotted ellipse in (b); compare with Fig.~\ref{Fig_harmonic}. The system parameters are $V_0\!=\!0.25 t_{\text{NN}}$, $t_{\text{NNN}}\!=\!0.4 t_{\text{NN}}$, $R_{\text{inter}}\!=\!17 a$, and $L_x\!=\!L_y\!=\!75a$. In panel (a), the trap $V_{\text{trap}} (\bs x)$ is measured in units of $t_{\text{NN}}$.}\label{Fig_harmonic_radial}\end{figure}

%\newpage
%\newpage
\section{Wave-packet dynamics}\label{Section:dynamics}

In this section, we explore the dynamical properties of our topological-interface scheme, by studying the motion of Gaussian wave packets in various relevant configurations. 
The wave packets are defined as
\begin{align}
&\vert \psi_{\text{0}} \rangle = \sum_{j} G_j \hat a_j^{\dagger} \vert 0 \rangle, \\
&G_j = (1/\mathcal{N}) \e^{-(x_j - x_0)^2/2\sigma_x^2}\e^{-(y_j - y_0)^2/2\sigma_y^2}\e^{i k_x^0(x_j - x_0)}\e^{i k_y^0(y_j - y_0)},\nonumber
\end{align}
The time-evolution of wave packets presented below was obtained through a numerical implementation of the time-evolution operator $\exp (-i t \hat H/\hbar)$ associated with the full 2D system, including the effects of the external trap introduced above, and the space-dependent offset $\DAB(\bs x)$. This study aims to highlight the applicability of our topological-interface scheme, in view of detecting the chiral motion of topological modes within an optical-lattice setup.

\subsection{The linear-interface case}\label{section:dynamics_linear}

Let us start by considering the linear-interface scheme associated with the space-dependent offset in Eq.~\eqref{space_offset}. We show in Fig.~\ref{Fig_dynamics_linear} the time-evolution of a small wave packet moving in the 2D lattice, for four different initial conditions. In the first case [Fig.~\ref{Fig_dynamics_linear}(a)], the wave packet is initially prepared on the central interface at $x\!=\!x_R$, with a mean quasi-momentum $\bs k^0\!=\!(0,2.1/a)$, which maximizes the projection unto the chiral localized mode [see the dispersion in Fig.~\ref{Fig_4}(a)]. We find that this initial state, whose small mean-deviation $\sigma_x\!=\!3a$ is of the order of the localization length of the central chiral mode, projects unto this localized mode with about $90\%$ efficiency. This wave packet undergoes a chiral motion along the topological interface, with positive mean velocity $v_g^y\!\approx\! 1.7 a t_{\text{NN}}/\hbar$, which is in agreement with the approximately-linear dispersion shown in Figs.~\ref{Fig_4}(a)-\ref{Fig_5}(b). We then show in Figs.~\ref{Fig_dynamics_linear}(b)-(c) how shifting the initial position of the wave packet, or changing its mean quasi-momentum, dramatically affects the projection unto chiral modes: In both these cases, the wave packet projects unto delocalized modes [with about $100\%$ efficiency], and accordingly, it undergoes an irregular (non-chiral) motion within the 2D lattice, and  diffuses into the bulk. Figure~\ref{Fig_dynamics_linear}(d) shows the dynamics of a wave packet initially prepared on the other interface at $x\!=\!x_L$, with a mean quasi-momentum $\bs k^0\!=\!(0,-2.1/a)$ that maximizes the projection unto the other localized mode: In this situation, the wave packet performs a regular motion along the second topological interface, $x\!=\!x_L$, with an opposite chirality. The center-of-mass motion of these wave packets, along the ``propagation" direction $y$, is further illustrated in Fig.~\ref{Fig_COM_linear}. As a technical remark, we note that the group velocity of the localized modes  slightly differ on the two interfaces $x_{R,L}$, which is due to the presence of the trap and microscopic details of the two interfaces.

This numerical study demonstrates how the topological chiral modes, which are localized on the topological interfaces, can be populated and probed \emph{in situ}, through a careful preparation of the wave-packet's position and momentum. We verified, through numerical simulations, that this preparation can be achieved by performing a partial Bloch oscillation: This consists in applying a force $\bs F\!=\! F_y \bs{1}_y$ for a  time $\Delta_t$, in such a way that the mean quasi-momentum $k^0_y(\Delta_t)\!=\! F \Delta_t$ reaches the desired value. Note that in this scheme, the initial (spatial) position of the wave packet should also be adjusted, so that it exactly reaches the desired topological interface after the duration $\Delta_t$.

\begin{figure}%[h!]
\includegraphics[width=9.3cm]{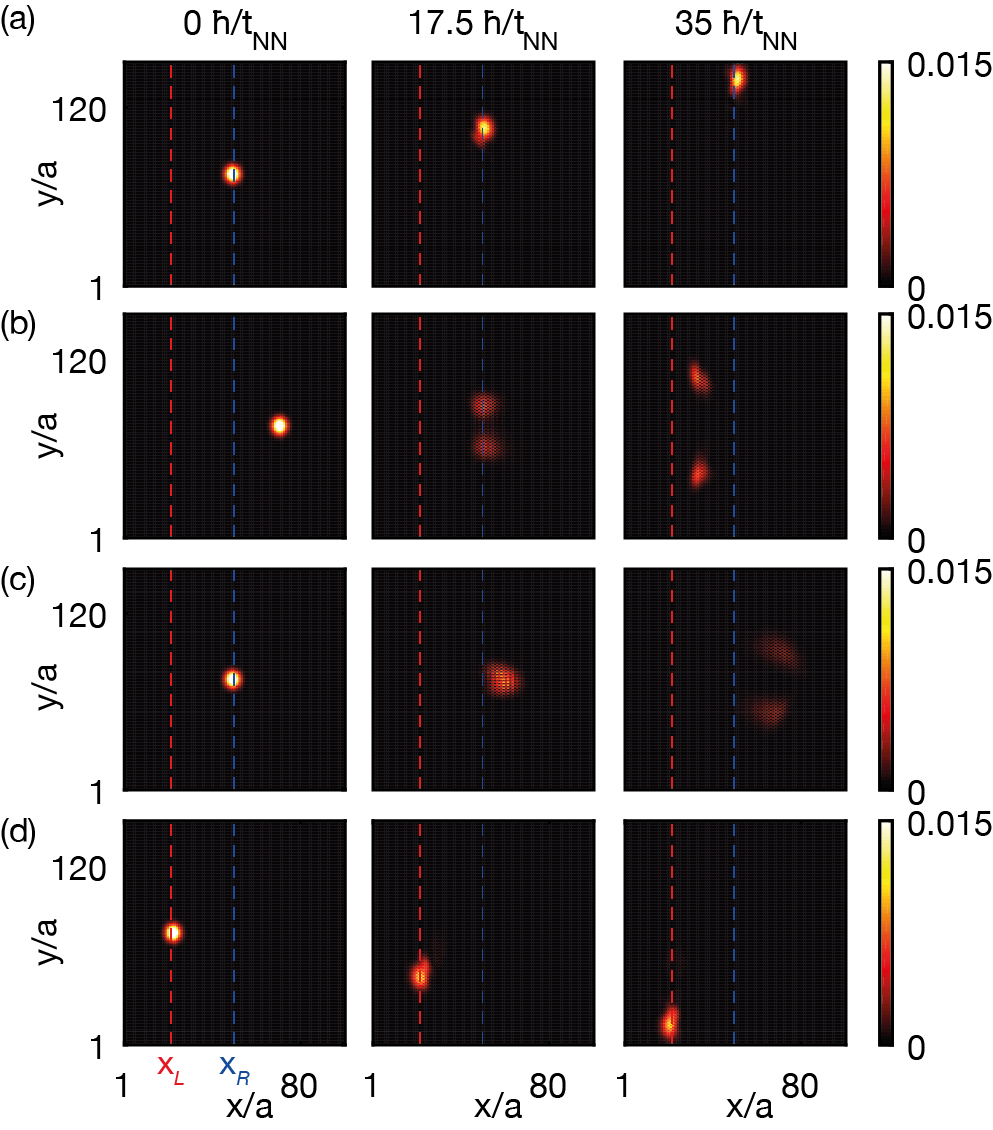}
\vspace{-0.cm} \caption{Wave-packet dynamics in the 2D honeycomb lattice. (a) A small Gaussian wave packet is initially prepared at the center of the system [i.e. on the central interface at $x\!=\!x_R$], and its mean quasi-momentum $\bs k^0\!=\!(0,2.1/a)$ is adjusted so as to maximize the projection unto the chiral localized modes. Due to its small mean-deviation $\sigma_x\!=\!3a$, which is of the order of the localization length of the central chiral mode, this wave packet projects unto this chiral mode with about $90\%$ efficiency, leading to a clear chiral motion along the central topological interface. (b) Same wave packet, but shifting its initial mean position away from the interface $x^0\!=\!x_R\!+\!20a$: This wave packet mainly projects unto the (non-chiral) delocalized states and diffuses into the bulk. (c) The wave packet is prepared at the center of the system, but with an opposite initial mean quasi-momentum $\bs k^0\!=\!(0,-2.1/a)$: As in (b), the wave packet projects unto delocalized states and diffuses into the bulk. (d) Initially preparing the wave packet on the other interface [$x\!=\!x_L$], with mean quasi-momentum $\bs k^0\!=\!(0,-2.1/a)$: This wave packet projects unto the other localized mode and undergoes the opposite chiral motion. In all cases, the system parameters are $t_{\text{NNN}}\!=\!0.4 t_{\text{NN}}$, $\delta_{\text{max}}\!=\!20 t_{\text{NN}}$, $L_x\!=\!100a$ and $L_y\!=\!150a$, and the trap parameters are $V_{x}\!=\!V_{y}\!=\!0.2 t_{\text{NN}}$ and $L_{\text{obs}}\!=\! (L_y/2) \sqrt{V_y/t_{\text{NN}}}$. In all figures, time is expressed in units of $\hbar/t_{\text{NN}}$.}
\label{Fig_dynamics_linear}
\end{figure}

\begin{figure}%[h!]
\includegraphics[width=9.3cm]{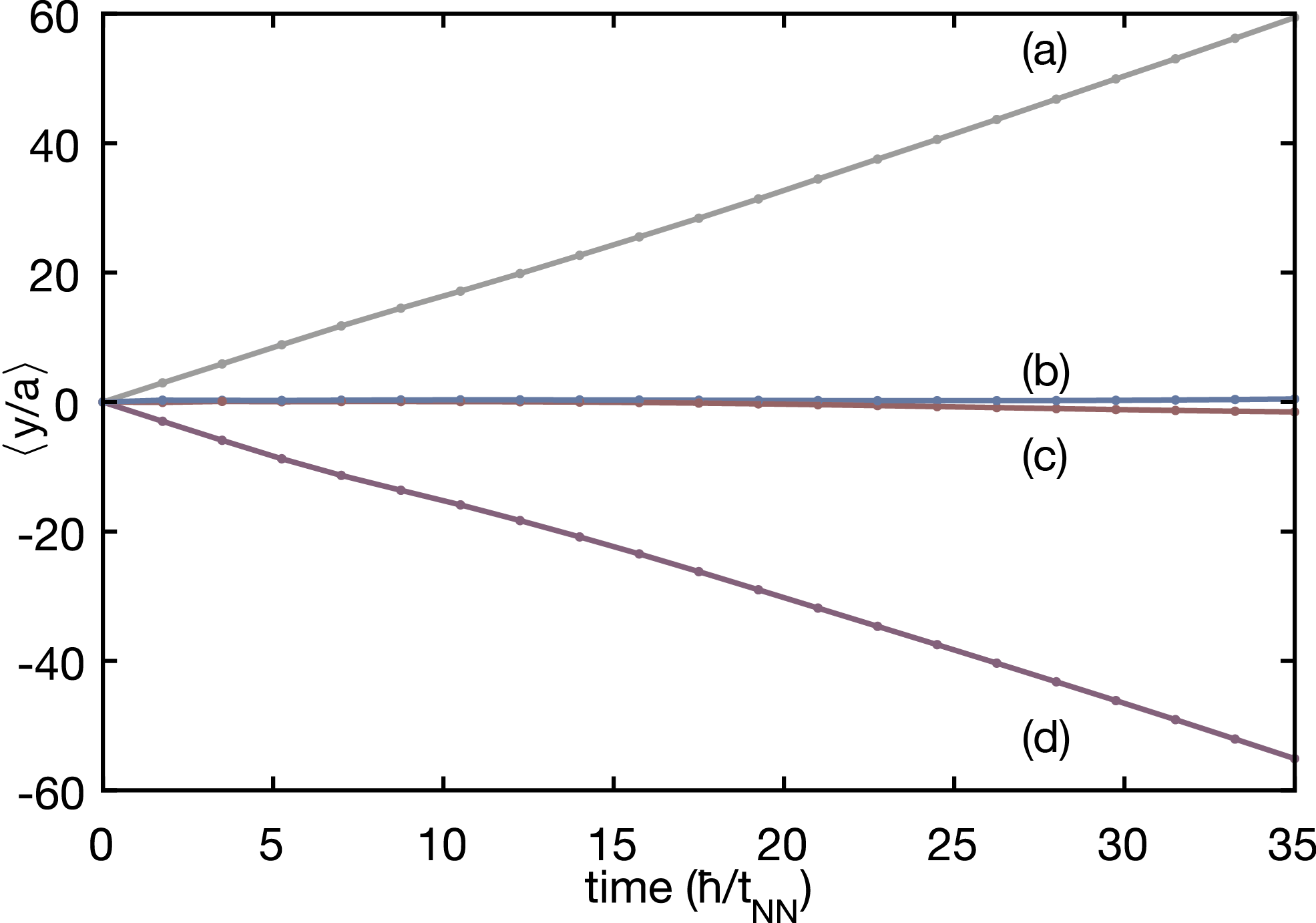}
\vspace{-0.cm} \caption{Center-of-mass evolution along the $y$ direction, $\langle y \rangle\!=\!\langle y \rangle(t)$, for the four situations [(a)-(d)] illustrated in Fig.~\ref{Fig_dynamics_linear}. The cases (a) and (d) correspond to opposite chiral motion along the interfaces $x_R$ and $x_L$, respectively. The cases (b) and (c) illustrate non-chiral motion generated by the bulk (delocalized) states. Time is expressed in units of $\hbar/t_{\text{NN}}$.}
\label{Fig_COM_linear}
\end{figure}

\subsection{Large clouds and the differential measurement}

In the previous Section~\ref{section:dynamics_linear}, we investigated the dynamics of small wave packets, whose size maximizes the projection unto the localized mode; in the realistic system configurations considered here, the localization length is found to be of the order of five lattice sites [see also Figs.~\ref{Fig_4} and \ref{Fig_2D_states}(b)]. In this section, we discuss the fate of more experimentally realistic wave packets, whose size typically exceeds the localization length of the topological modes. 

\subsubsection{Dynamics of larger clouds}\label{Section:large}

The dynamics shown in Fig.~\ref{Fig_larger_dynamics}(a) corresponds to the same system configuration as in Fig.~\ref{Fig_dynamics_linear}(a), but with an initial Gaussian wave packet of mean-deviation $\sigma_x\!=\!20 a$ instead of $\sigma_x\!=\!3 a$ (and a slightly larger system size along the propagation direction $y$). Due to its larger width, this initial state only projects with about $30\%$ efficiency unto the chiral mode that is localized on the central topological interface. This leads to significant bulk noise in the background of the moving cloud, as compared with the result shown in Fig.~\ref{Fig_dynamics_linear} (a), which limits the detection of the chiral motion associated with the localized topological mode. We emphasize that, although the larger cloud potentially overlaps with the other topological interface at $x\!=\!x_L$, this initial state does not project unto the corresponding mode (of opposite chirality): This is due to the fact that the dispersion relations of the two counter-propagating modes are associated with disconnected regions in $\bs k$-space [Fig.~\ref{Fig_4}(a)]. Note that in experiments, in particular when using fermionic atoms, the cloud may be broad in both real- and momentum-space, which is expected to increase the contribution of bulk states.

\subsubsection{The differential measurement}\label{section_differential}

As proposed in Ref.~\cite{GoldmanPNAS}, a differential measurement can be performed to improve the detection of chiral modes in the presence of noisy backgrounds, which are typically associated with the  contribution of delocalized states to the particle density. This idea is based on the fact that only chiral modes are severely affected when performing a time-reversal (TR) transformation to the system, as we now briefly recall. For a 2D square lattice in a uniform magnetic field, this TR transformation consists in reversing the sign of the applied magnetic field: This leaves the dispersion of the bulk bands perfectly unchanged, but it reverses their Chern number, and hence, also the chirality of all the topological edge modes present in the bulk gaps \cite{GoldmanPNAS}: Hence, subtracting the particle density associated with two time-reversal-related configurations potentially annihilates any contribution from the bulk, allowing for clean detection of the chiral modes dynamics. In the context of the Haldane model, Eq.~\eqref{Eq:Haldane}, this transformation consists in reversing the sign of the TR-breaking term, i.e.~$t_{\text{NNN}}\!\rightarrow\! -t_{\text{NNN}}$. As a technical remark, one should note that in the presence of the offset $\DAB$, inversion symmetry is also broken: As a consequence, the bulk dispersions are no longer perfectly immune to the TR transformation. This observation is irrelevant when considering a completely filled band, but it does affect the differential measurement when it is applied to wave packets that are localized in $\bs k$-space, see e.g. Ref. \cite{Jotzu2014}.

We now illustrate how this differential measurement can be exploited in the present proposal, by applying it to the situation discussed above in Section~\ref{Section:large} and depicted in Fig.~\ref{Fig_larger_dynamics}(a).
First, in Fig.~\ref{Fig_larger_dynamics}(b) we show the time-reversal counterpart of Fig.~\ref{Fig_larger_dynamics}(a), which was obtained by reversing the sign of the TR-breaking term, i.e.~$t_{\text{NNN}}\!\rightarrow\! -t_{\text{NNN}}$, as well as the sign of the mean quasi-momentum $\bs k^0\!\rightarrow\! -\bs k^0$ of the initial Gaussian wave packet. 
Even if the particle density again shows a significant contribution from the delocalized states, this result shows how the TR transformation indeed reverses the general direction of propagation along $y$. Next, we subtract the densities associated with the two TR-counterparts, and we show the corresponding differential measurement in Fig.~\ref{Fig_larger_dynamics}(c). As anticipated above, we find that the contribution of the delocalized (bulk) states is largely annihilated by the differential measurement, allowing for a clear detection of the chiral modes (including the measurement of their group velocity), even in the regime of large atomic clouds. We note that the residual noise, which is visible in Fig.~\ref{Fig_larger_dynamics}(c), is due to a weak asymmetry in the bulk bands, which is a direct consequence of the inversion-symmetry-breaking offset $\DAB$, as announced above.

\begin{figure}%[h!]
\includegraphics[width=9.3cm]{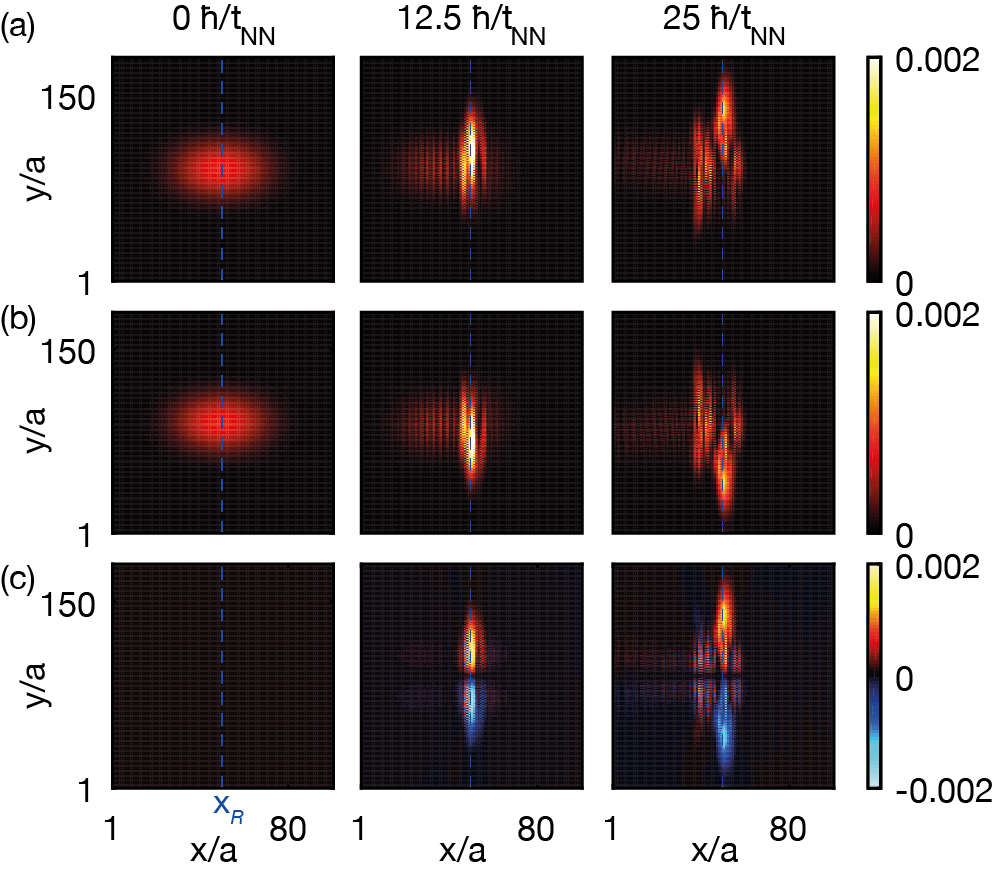}
\vspace{-0.cm} \caption{Dynamics in the 2D honeycomb lattice for a large wave packet, of mean deviation $\sigma_x\!=\!20 a$. The system parameters are the same as in Fig.~\ref{Fig_dynamics_linear}, except $L_y\!=\!180a$. (a) The large wave packet is prepared on the central topological interface, with mean quasi-momentum $\bs k^0\!=\!(0,2.1/a)$, so as to maximize the projection unto the localized modes [as in Fig.~\ref{Fig_dynamics_linear} (a)]. However, due to its large width, this state only projects unto the localized mode with about $30\%$ efficiency, which leads to significant bulk noise in the background of the moving cloud [compare with Fig.~\ref{Fig_dynamics_linear} (a)]. (b) The time-reversal (TR) counterpart of the configuration depicted in (a), namely, when reversing the TR-breaking term $t_{\text{NNN}}\!\rightarrow\! -t_{\text{NNN}}$ and the mean quasi-momentum of the initial state $\bs k^0\!\rightarrow\! -\bs k^0$. (c) The differential measurement, obtained by subtracting the TR-related situations in (a) and (b), allows one to significantly reduce the noise associated with the bulk contributions, and hence, highlight the motion of topological chiral modes. In all figures, time is expressed in units of $\hbar/t_{\text{NN}}$.}
\label{Fig_larger_dynamics}
\end{figure}

\subsection{The radial-interface case}

In this section, we present the time-evolution of a wave packet initially prepared on the radial topological interface defined in Eq.~\eqref{space_offset_radial}, and in the presence of the trap [Eq.~\ref{trap}]. The corresponding time-evolving particle density is shown in Fig.~\ref{Fig_radial_dynamics}, for a Gaussian wave packet whose mean-deviation and phase have been adjusted so as to maximize projection unto the chiral localized mode. This result highlights the advantage of probing the physics of chiral modes using radial topological interfaces, as the latter is particularly immune to the presence of the harmonic trap [see also Fig.~\ref{Fig_harmonic_radial}], and hence allows for the analysis of edge-state physics over ``arbitrarily" long observation times. 

\begin{figure}[h!]
\includegraphics[width=9.3cm]{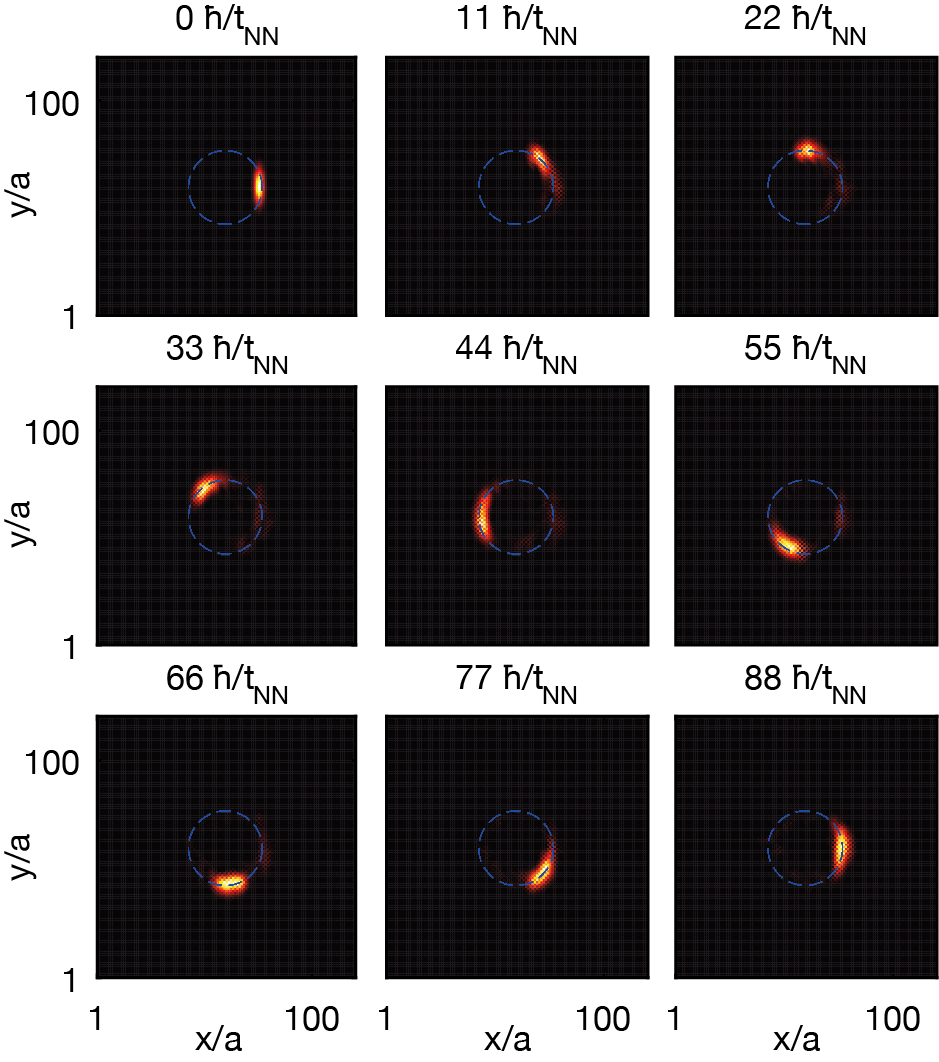}
\vspace{-0.cm} \caption{Wave-packet dynamics in the 2D honeycomb lattice for the radial-interface configuration [Eq.~\eqref{space_offset_radial}], with an interface radius $R_{\text{inter}}\!=\!17a$, $t_{\text{NNN}}\!=\!0.4 t_{\text{NN}}$, and a harmonic potential of strength $V_0\!=\!0.25t_{\text{NN}}$, see also Eq.~\ref{trap} and Fig.~\ref{Fig_harmonic_radial}. In all figures, time is expressed in units of $\hbar/t_{\text{NN}}$.}
\label{Fig_radial_dynamics}
\end{figure}

\section{Studying the topological nature of interfaces using disorder}\label{Section:disorderb}

The observation of the quantum Hall effect in solid-state physics relies on the robustness of chiral edge modes against disorder~\cite{Yoshioka}. Indeed, whereas the bulk states are localized by disorder, the chiral nature of topological edge modes prevents them from any backscattering processes (as long as opposite edges are well spatially separated). In this section, we verify that the topological modes that propagate along engineered topological interfaces within the system are equally robust against disorder. In our study of disorder, we take this perturbation to be in the form of a random (on-site) potential, with energies uniformly distributed between 0 and $D\!=\!2 t_{\text{NN}}$. We point out that such a disordered potential can be engineered in a cold-atom experiment, using optical speckle potentials \cite{Sanchez-Palencia/Lewenstein}. 

In Fig.~\ref{Fig_small_dynamics_disorder} we show how the dynamics of small wave packets is modified by disorder. Comparing these results with the clean situation previously shown in Figs.~\ref{Fig_dynamics_linear}(a)-(b) clearly reveals the robustness of the edge mode that propagates along the central topological interface [Fig.~\ref{Fig_small_dynamics_disorder} (a)], as well as the disorder-induced localization of wave packets made of bulk (non-chiral) states [Fig.~\ref{Fig_small_dynamics_disorder} (b)].

Most importantly, we show in Fig.~\ref{Fig_larger_dynamics_disorder}(a) how disorder can be \emph{exploited} to improve the detection of topological modes in cold-atom systems: By annihilating the dispersion of the bulk states, which typically constitute the majority of populated states in realistic situations involving wide atomic clouds, the disorder naturally enhances the signal associated with the propagating chiral modes. Indeed, the time-evolving cloud depicted in Fig.~\ref{Fig_larger_dynamics_disorder}(a) directly reveals the propagation of the central topological mode, in contrast with the disorder-free situation previously shown in Fig.~\ref{Fig_larger_dynamics} (a) for the same system parameters. Finally, Fig.~\ref{Fig_larger_dynamics_disorder}(b) shows how combining disorder and the aforementioned differential measurement [Section~\ref{section_differential}] allows one to reach a remarkably precise visualization of the edge-mode propagation, in the absence of residual noise associated with the bulk [compare with the disorder-free case in Fig.~\ref{Fig_larger_dynamics}(c)]. The fact that the differential measurement is improved by disorder relies on that this perturbation dephases the cloud: This smoothes out the residual background noise associated with the asymmetry in the bulk bands under the TR transformation [see the discussion in Section~\ref{section_differential}].

The results in Fig.~\ref{Fig_larger_dynamics_disorder} highlight how engineered disorder,  as created by optical speckle potentials \cite{Sanchez-Palencia/Lewenstein}, could be used as a powerful tool for the detection and study of topological-edge-state physics in cold-atom systems. In this sense, the important role played by disorder in revealing QH physics~\cite{Yoshioka} is thus not restricted to solid-state physics.

%\subsubsection{Robustness of the edge modes}

\begin{figure}%[h!]
\includegraphics[width=9.4cm]{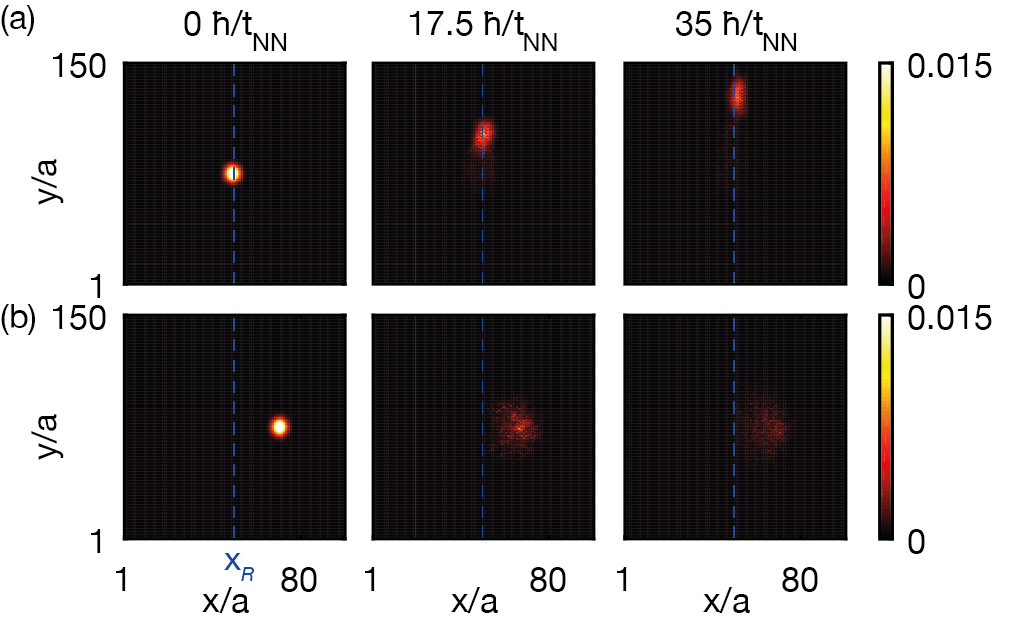}
\vspace{-0.cm} \caption{Wave-packet dynamics in the 2D honeycomb lattice for the same configuration as in Fig.~\ref{Fig_dynamics_linear}(a)-(b), but including a disordered potential of strength $D\!=\!2 t_{\text{NN}}$, and averaging over ten realizations of disorder. (a) The edge mode localized on the central topological interface is immune to disorder, and freely propagates along the topological interface. (b) A wave packet associated with (non-chiral) bulk states is localized by the disorder [compare with Fig.~\ref{Fig_dynamics_linear}(b)].}
\label{Fig_small_dynamics_disorder}
\end{figure}

%\subsubsection{Using disorder to improve edge-mode detection}

\begin{figure}%[h!]
\includegraphics[width=9.3cm]{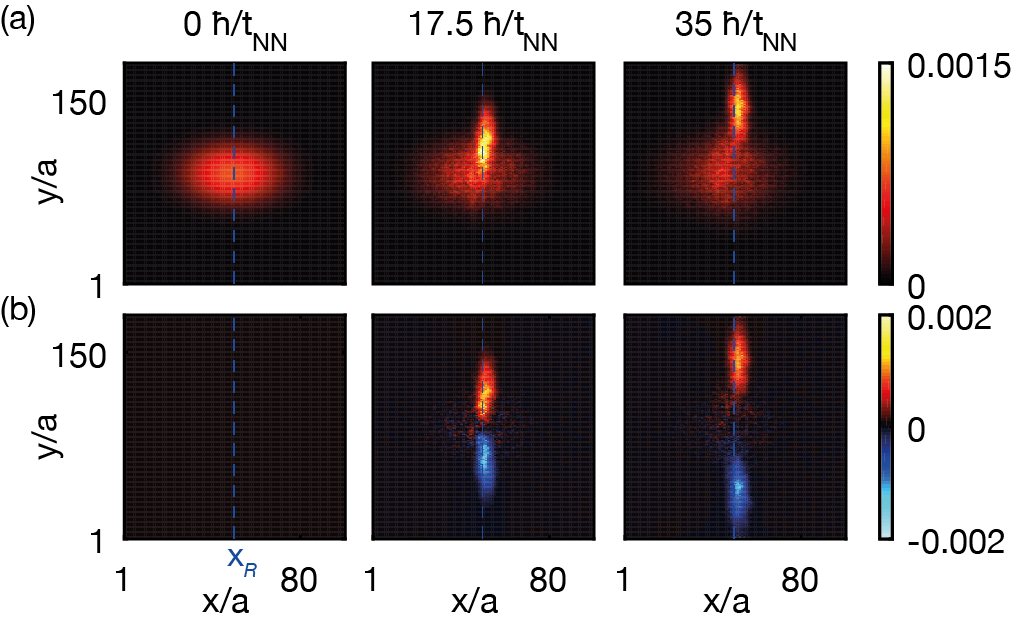}
\vspace{-0.cm} \caption{Dynamics of a large wave packet, in the same configuration as in Fig.~\ref{Fig_larger_dynamics}(a), but including a disordered potential of strength $D\!=\!2 t_{\text{NN}}$, and averaging over thirty realizations. (a) The chiral motion of the localized mode is clearly identified in the presence of disorder, as the latter annihilates the propagation of the bulk states. (b) The differential measurement in the presence of disorder reveals clean edge-mode dynamics at the central topological interface, by removing the contribution of the (dephased) bulk states to the density [compare with Fig.~\ref{Fig_larger_dynamics}(c)].}
\label{Fig_larger_dynamics_disorder}
\end{figure}

\section{Experimental implementation}\label{Section:experiment}
\subsection{The linear interface}\label{Section:experiment_lin}

\begin{figure}%[h!]
\includegraphics[width=8.6cm]{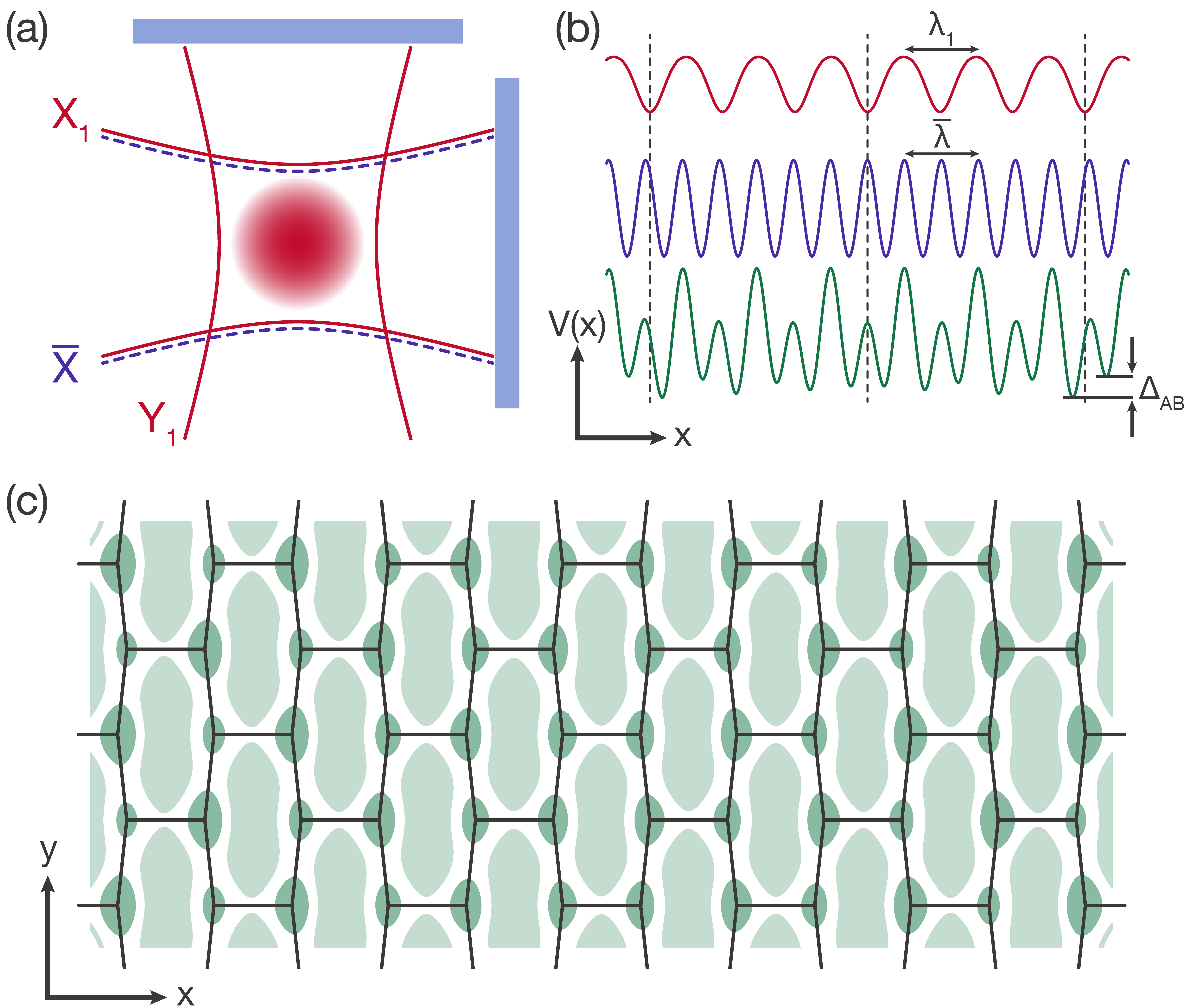}
\vspace{-0.cm} \caption{Experimental implementation of the linear scheme. 
(a) Beam setup. The retro-reflected laser beams $X_1$ and $Y_1$, with wavelength $\lambda_1$ interfere with each other at the position of the atomic cloud (red) and are phase stabilized. The additional beam $\overline{X}$ (blue) has a different wavelength $\overline{\lambda}$ and has the same transverse spatial mode as $X_1$. 
(b) Cut through the optical lattice potentials at $y=0$. The standing wave created by $\overline{X}$ (blue) has a lattice spacing of $\overline{\lambda}/2$, which is slightly more than half of $\lambda_1$, the $x$-direction spacing of the interference lattice created by $X_1$ and $Y_1$ (red). Therefore, their extrema only coincide at a particular point in the system. The resulting total potential (green) then has a spatially varying site-offset $\DAB$.
(c) Sketch of the resulting potential, with the tight-binding lattice structure (black) superimposed. Minima are dark green, maxima light green. The variation of $\DAB$ in the $x$ direction is exaggerated for better visibility.}
\label{Fig_Setup1}
\end{figure}

The honeycomb lattices with a spatially dependent site offset discussed in this work can be implemented experimentally using an extension of the tunable-geometry lattice introduced in Ref.~\cite{Tarruell2012}. 
A linear variation of $\DAB$, as introduced in Eq.~\eqref{space_offset}, can be created as follows:
First, a pair of red-detuned retro-reflected laser beams with identical wavelength $\lambda_1$ and single-beam lattice depths $V_{X1}$ and $V_{Y1}$ are phase-stabilized with respect to each other and oriented along the $x$ and $y$ direction respectively [see Fig. \ref{Fig_Setup1}(a)]. 
At their intersection, where the atomic cloud is placed, the resulting potential experienced by the atoms is given by 
\begin{align}
V_1(x,y) = & - V_{X1} \cos^2(k_1
x)-V_{Y1} \cos^2(k_1 y)\nonumber \\
&- 2\sqrt{V_{X1}V_{Y1}}\cos(k_1 x)\cos(k_1
y), \label{eqlattice_xy1}
\end{align} 
where $k_1=2\pi/\lambda_1$, which corresponds to a checkerboard lattice 
\footnote{Here and henceforth, the potential is given in the $xy$ plane only. In the $z$ direction, either a weak harmonic trap, as in Ref.~\cite{Tarruell2012}, or an additional optical lattice, as in Ref.~\cite{Uehlinger2013PRL}, can be used. }.
Its unit vectors are oriented at $\pm 45^{\circ}$ with respect to the $x$ axis and the site spacing along the $x$ direction is $\lambda_1$, see Fig.~\ref{Fig_Setup1}(b).

An additional laser beam with lattice depth $V_{\overline{X}}$, operating at wavelength $\overline{\lambda}$ and oriented along the $x$ direction, gives rise to an additional standing wave with site spacing $\overline{\lambda}/2$ and potential 
\be
\overline{V}(x,y) = -V_{\overline{X}} \cos^2(\overline{k}x+\theta/2),
\label{eqlattice_xb}
\ee
where $\overline{k}=2\pi/\overline{\lambda}$.
We first consider the case where $\overline{\lambda}$ is so close to $\lambda_1$ 
that we can assume $\overline{k}=k_1$ over the size of the atomic cloud. 
However, because of the distance between the retro-reflecting mirror and the cloud, the small difference between $\overline{\lambda}$ and $\lambda_1$ (typically on the order of $1$\,pm or - in terms of frequency - a few hundred MHz) still leads to a shift between the two potentials, which is captured \textit{via} $\theta$.  
Setting $\theta\approx\pi$, this gives rise to the honeycomb lattice of Ref. \cite{Tarruell2012}. 
Its near-constant site-offset $\DAB$ is tuned via $\theta$ and can be calibrated using Bloch-Zener oscillations \cite{Uehlinger2013EPJ}. It becomes 0 when $\theta\!=\!\pi$. 

In order to achieve a significant spatial variation of $\DAB$ over the size of the cloud, we increase the difference between $\overline{\lambda}$ and $\lambda_1$, such that $\overline{k}=k_1$ is not a good approximation any more. 
Then, the extrema of $\overline{V}(x,y)$ and $V_1(x,y)$ will line up differently depending on the position within the atomic cloud, leading to a smooth variation of $\DAB$ along the $x$ direction, as illustrated in Fig. \ref{Fig_Setup1}(b)-(c). 

The tight-binding parameters corresponding to this optical lattice depend on the choice of atomic species and laser wavelength \cite{Lewenstein2012}.
For example, when using $^{40}$K atoms and setting the laser intensity such that $t_{\mathrm{NN}}=h\times 200$\,Hz, a variation of $0.2t_{\mathrm{NN}}$ per site (i.e. $\delta_{\text{max}}/L_x=0.2t_{\mathrm{NN}}/a$ in Eq.~\eqref{space_offset} is achieved with $\lambda_1 = 1064.0$\,nm and $\overline{\lambda} = 1064.5$\,nm
\footnote{
For comparison, in the case discussed above, the ``near-constant'' $\DAB$ as realized in Ref.~\cite{Tarruell2012}, typically corresponds to changes in $\DAB$ of about $5 \times 10^{-4}t_{\mathrm{NN}}$ per site.
}.
For these parameters, the spatial variation of $\DAB$ deviates from a linear function by less than $0.1\%$, whilst $t_{\mathrm{NN}}$ changes by at most 1\% over a range of 100 sites.

\subsection{The radial-symmetric scheme}\label{Section:experiment_circ}

\begin{figure}%[h!]
\includegraphics[width=8.6cm]{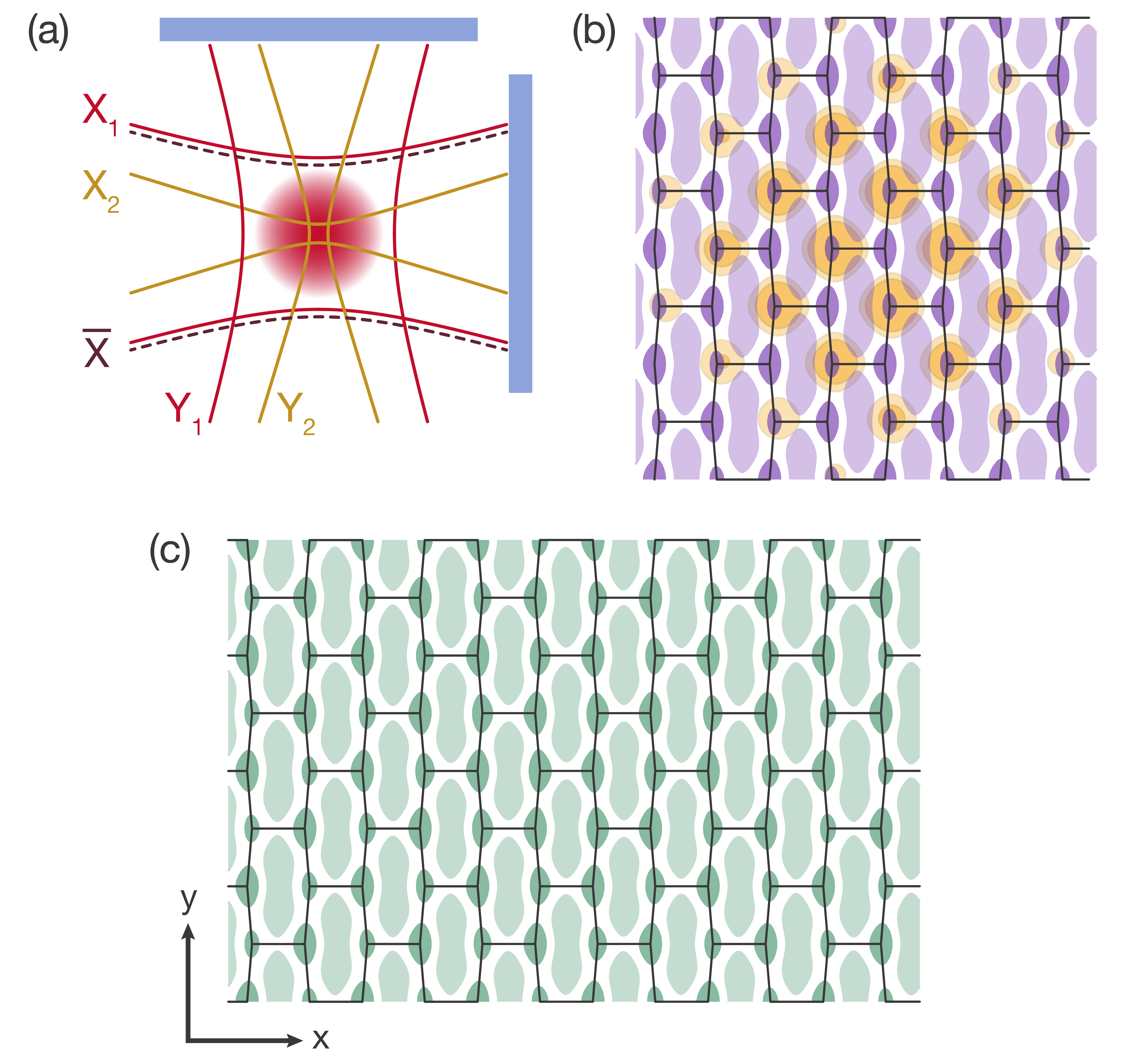}
\vspace{-0.cm} \caption{
Experimental implementation of the radial-symmetric scheme. 
(a) Beam setup. Lasers $X_1$, $Y_1$ (red) and $\overline{X}$ (dark red, dashed) are similar as in Fig.~\ref{Fig_Setup2}, except that now $\overline{\lambda} \approx \lambda_1$. In addition, a pair of phase-stabilised beams $X_2$ and $Y_2$ with a smaller beam waist are focused onto the center of the atomic cloud.
(b) Laser beams $X_1$ and $Y_1$, together with $\overline{X}$ form an imbalanced honeycomb lattice potential, which is nearly uniform over the extent of the atomic cloud (dark blue minima, light blue maxima, tight-binding structure indicated by black lines). 
The minima of the additional potential (orange) created by $X_2$ and $Y_2$ are aligned with every second site of the honeycomb lattice. Their intensity decreases away from the center.
(c) The resulting potential has a radially varying site offset (dark green minima, light green maxima). The spatial variation is exaggerated for better visibility.}
\label{Fig_Setup2}
\end{figure}

In order to create a two-dimensional radial-symmetric variation of $\DAB$, as introduced in Section \ref{section:radial_interface}, a different scheme can be used: 
A honeycomb lattice with a near-constant site offset is created by laser beams $X_1, Y_1$ and $\overline{X}$ (taking $\overline{\lambda}$ very close to $\lambda_1$ 
so that $k_1 = \overline{k}$ is a good approximation) as outlined above and shown in Fig.~\ref{Fig_Setup2}(a)-(b). Its lattice structure can be assumed homogeneous over the size of the atomic cloud. 

Two additional phase-stabilised, retro-reflected laser beams $X_2$ and $Y_2$, operating at $\lambda_2 \approx \lambda_1$ have much smaller transverse beam waists, $w$, than the other beams. The checkerboard lattice potential they create is given by 
\begin{align}
V_2(x,y)  = &  \;\e^{-2(x^2+y^2)/w^2}\times \nonumber \\ 
 \big( &-V_{X2} \cos^2(k_2
x+\theta/2)-V_{Y2} \cos^2(k_2 y)\nonumber \\
&- 2\sqrt{V_{X2}V_{Y2}}\cos(k_2 x + \theta/2)\cos(k_2y) \big). 
\label{eqlattice_xy2}
\end{align} 
The small detunings between $\lambda_2$, $\lambda_1$ and $\overline{\lambda}$ as well as the distances between retro-reflecting mirrors and the atomic cloud are chosen such that $\theta$ is the same as in Eq.~\eqref{eqlattice_xb} and we can assume $k_2=k_1=\overline{k}$. Then, the minima of this potential coincide with every other site (i.e. with either the A or B sublattice, see Fig.~\ref{Fig_one}) of the honeycomb lattice [see Fig.~\ref{Fig_Setup2}(b)], thereby contributing to the site offset $\DAB$. The Gaussian envelope of the beams means that this contribution varies spatially. For example, by choosing $V_{X2}$ and $V_{Y2}$ correctly, we can set $\Delta_{\mathrm{AB}}=0$ in the center of the cloud and let it increase as a Gaussian function of the radial distance to the center, as illustrated in Fig.~\ref{Fig_Setup2}(c).

\section{Concluding remarks and outlooks}\label{Section:conclusions}

In this work, we introduced a novel scheme allowing for the direct detection of topological propagating modes within 2D ultracold atomic gases. Our proposal is based on the engineering of topological interfaces, which localize topologically-protected  modes in desired regions of space (e.g.~at the center of a harmonic trap, typically present in cold-atom experiments). This allows for real-space detection of topological transport in a highly controllable and versatile platform. In particular, we stress that the trajectory performed by these topologically-protected modes within the system can be tuned by shaping the form of the topological interfaces, which suggests interesting applications based on topological quantum transport. In particular, this opens an exciting avenue for the manipulation and probing of topological modes in atomic systems, where disorder, inter-particle interactions and external gauge fields can be induced and controlled at will. We note that similar topological interfaces could be created within photonic crystals, where system parameters (e.g.~the hopping amplitudes and on-site potentials entering engineered tight-binding models) can also be precisely addressed in a local manner~\cite{Longhi:2006,Mukherjee:2016}. 

%The results presented in this article focus on the edge properties associated with engineered interfaces. However, we point out that the topology of the bulk, which varies across each interface, could equally be probed. This could be realized, for instance, by generalizing the Chern-number measurement reported in Ref.~\cite{Aidelsburger2015}: if the atomic cloud is prepared in some region $R_1$, where the lowest bulk band is associated with the Chern number $C_1$, and if the atoms uniformly fill this band (e.g.~through thermal effects or Fermi statistics), then the center-of-mass trajectory of the cloud in response to an applied force will directly reveal the local Chern number $C_1$ \cite{Dauphin:2013,Price:2016}. We note that interesting dynamics could be investigated in situations where the atomic cloud visits different topologically-ordered regions of space as it drifts in response to the applied force; besides, rich dynamics are also expected if the atomic cloud is initially prepared within different topologically-ordered regions of space (in which case the center-of-mass drift will be dictated by a series of Chern numbers).

The scheme introduced in this work has been illustrated based on a 2D model~\cite{Haldane1988}, which exhibits the integer quantum Hall effect; namely, a non-interacting system featuring 2D Bloch bands with non-zero Chern numbers~\cite{Nagaosa:2010}. However, we point out that our scheme, which consists in varying a model parameter in space in view of creating local topological phase transitions (i.e.~topological interfaces), can be applied to any model exhibiting topological band structures. For instance, similar interfaces could be engineered in the quantum-spin-Hall regime of $Z_2$ (time-reversal-invariant) topological insulators: Considering the 2D Kane-Mele model~\cite{Kane:2005}, which is a direct extension of the Haldane model~\cite{Haldane1988} to spin-1/2 particles, this could be realized either by introducing a spatially-varying offset $\Delta_{\mathrm{AB}}$ between the sites of the honeycomb lattice (as proposed in this work), or by engineering a spatially-varying spin-orbit coupling. In this spinful TRS situation, the topological interfaces would host helical topological modes, namely, modes associated with opposite spins and propagating in opposite directions (see Refs.~\cite{Kane:2005,GoldmanPRL2010} and the recent photonics proposal~\cite{Barik:2016}). We stress that, in contrast to the spinless TRS configuration of Section~\ref{Section:TRS}, these helical counter-propagating modes are topologically protected (backscattering processes are forbidden by time-reversal symmetry in this spinful case~\cite{Kane:2005}). Hence, engineering  interfaces in $Z_2$ topological insulators would introduce adjustable guides for topologically-protected spin transport within 2D systems. The same strategy could be applied to higher-dimensional systems, such as 3D topological insulators~\cite{Fu:2007}: By varying the spin-orbit coupling of such systems in space, one could engineer 2D topological interfaces hosting a single 2D Dirac fermion. In this scheme, these intriguing excitations would be located \emph{within} the system, instead of at its surfaces, which could also open interesting avenues for spin transport in 3D systems. It is worth pointing out that $Z_2$ topological insulators can be realized in 2D optical lattices~\cite{GoldmanPRL2010,Beri:2011,Kennedy:2013}; see Ref.~\cite{Aidelsburger2013} for a first experimental realization of such a model with cold atoms. Furthermore, these setups could be extended in view of creating 3D topological insulators~\cite{Bermudez:2010}, but also, to reveal the 4D QH effect~\cite{Price:2015,Price:2016}, in cold-atom experiments. Creating topological interfaces within a 4D QH atomic system~\cite{Price:2015} offers a unique platform to investigate 3D topological surface modes (i.e.~spatially isolated Weyl fermions) in the laboratory.

Finally,  the physics of topological edge modes is certainly not restricted to non-interacting quantum systems. In particular, edge modes play a crucial role in fractional quantum Hall (FQH) liquids, where they present exotic (sometimes counter-intuitive) structures~\cite{Wen:1995}. For instance, FQH liquids potentially exhibit counter-propagating edge modes (allowing back-scattering on the edge in the presence of impurities), while still presenting a finite and quantized Hall conductivity~\cite{Haldane:1995,Wen:1995,Kane:1995,Kane:1994,Inoue:2013}. While these counter-propagating modes remained undetectable in standard edge-magnetoplasma experiments~\cite{Ashoori:1992,Wen:1995}, they were recently revealed through shot noise~\cite{Bid:2010,Gurman:2012,Inoue:2013} and  thermometry~\cite{Venkatachalam:2012} measurements. Even more recently, it was suggested in Ref.~\cite{Goldstein:2016} that the presence of such counter-propagating edge modes could have important consequences on the detection of fractional (anyonic) statistics, based on QH Mach-Zehnder interferometers~\cite{Ji:2003}. The unique possibility of creating topological interfaces in a cold-atom experiment offers a promising platform for the analysis of these topological edge structures, where  Bragg spectroscopy~\cite{Liu:2010,GoldmanPRL} and high-resolution imaging techniques could be exploited in view of revealing their exotic dispersion relations and dynamical properties in the presence of controllable interactions and disorder.

\acknowledgements We acknowledge A. Dauphin, M. Kolodrubetz, and D. T. Tran for helpful discussions. N.G. is financed by the FRS-FNRS Belgium and by
the BSPO under PAI Project No. P7/18 DYGEST. The ETH team acknowledges SNF, NCCR-QSIT, and QUIC (Swiss State Secretary for Education, Research and Innovation contract number 15.0019) for funding. R.D. acknowledges support from ETH Zurich Postodoctoral Program and Marie Curie Actions for People COFUND program.

%\bibliography{References}

%%%%%%%%%%%%%%%%%%%%%%%%%%%%%%%%%%%%%%%%%%%%%%%%%%%%%%%%%%%%%%%%%%%


%%%%%%%%%%%%%%%%%%%%%%%%%%%%%%%%%%%%%%%%%%%%%%%%%%%%%%%%%%%%%%%%%%%

\newpage

\vspace{1cm}

\appendix

\section{Dynamics in the time-reversal-invariant case}\label{Appendix:TRS}

In this Appendix, we show the dynamics of a wave packet, which is prepared on the central interface [$x\!=\!x_R$] for the time-reversal-invariant case where $t_{\text{NNN}}\!=\!0$. As shown in Fig.~\ref{Fig_Appendix_TRS}, tuning the mean quasi-momentum $k_y^0$ of the wave packet allows one to select between the two different edge channels, with opposite chirality. In the time-reversal-invariant case, the interface is thus essentially equivalent to an isolated 1D lattice, whose position within the 2D lattice can be tuned by varying the space-dependent offset $\DAB(x)$.

\begin{figure}[h!]
\includegraphics[width=9.cm]{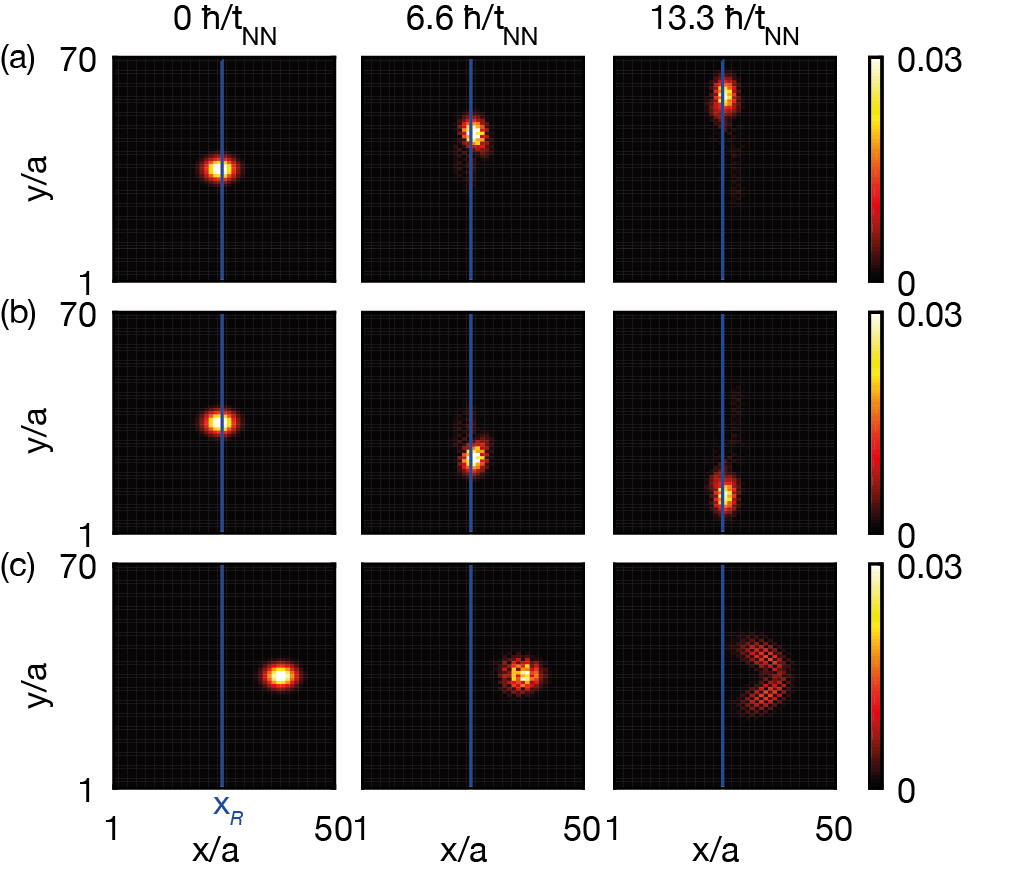}
\vspace{-0.cm} \caption{Wave-packet dynamics in the time-reversal case $t_{\text{NNN}}\!=\!0$, and $\delta_{\text{max}}\!=\!20 t_{\text{NN}}$. (a) By setting the mean position of the Gaussian wave packet on the interface ($x\!=\!x_R$), and the mean quasi-momentum along $y$ to be $k_y^0\!\approx\!+2.1/a$, we find that the Gaussian wave packet essentially projects unto the ``green" states with positive group velocity in Fig.~\ref{Fig_TRS}(b): The wave packet goes up along the central interface [$x\!=\!L_x/2$] where these states are localized. (b) By setting the mean quasi-momentum along $y$ to be opposite, $k_y^0\!\approx\!-2.1/a$, the wave packet essentially projects unto the states with negative group velocity, which are also localized at $x\!=\!L_x/2$: The wave packet goes down along the central interface. (c) Keeping $k_y^0\!\approx\!+2.1/a$, but now setting the mean position of the Gaussian wave packet away from the interface: We find that the wave-packet mainly projects unto standard (2D) bulk states, and hence, undergoes a non-chiral motion in the plane. Time is expressed in units of $\hbar/t_{\text{NN}}$.}\label{Fig_Appendix_TRS}\end{figure}

 \end{document}